\documentclass[useAMS,usenatbib,times]{mn2e}
\input{epsf.sty}

\def\lsim{\mathrel{\hbox{\rlap{\hbox{\lower4pt\hbox{$\sim$}}}\hbox{$<$}}}}
\def\etal  {\rm {et al.}\rm}

\newcommand{\sersic}[0]{S\'{e}rsic}

\newcommand{\bmgc}{B_{\mbox{\tiny \sc MGC}}}

\begin{document}

\title[MGC: Bulge/Disc Decomposition of 10095 Galaxies]
{The Millennium Galaxy Catalogue: Bulge/Disc Decomposition of 10095 Nearby Galaxies} 
\author[Allen \etal]
{
Paul D. Allen$^{1}$\thanks{paul@mso.anu.edu.au}, Simon P. Driver$^{1,2}$, Alister W. Graham$^{1}$, Ewan Cameron$^{1}$,
\newauthor
Jochen Liske$^{3}$, \& Roberto De Propris$^{4}$
\\
$^{1}$Research School of Astronomy and Astrophysics, The Australian National University, Mount Stromlo Observatory, Cotter Rd, \\  Weston, ACT 2611, Australia.\\
$^{2}$School of Physics and Astronomy, North Haugh, St Andrews, Fife, KY16 9SS, UK.\\
$^{3}$European Southern Observatory, Karl-Schwarzchild-Str. 2, D-85748, Garching bei M{\" u}nchen, Germany.\\
$^{4}$Cerro Tololo Inter-American Observatory, Casilla 603, La Serena, Chile.\\
}

\maketitle

\begin{abstract}
We have modelled the light distribution in 10095 galaxies from the Millennium 
Galaxy Catalogue (MGC), providing publically available structural 
catalogues for a large, representative sample of galaxies in the local 
Universe. Three different models were used: (1) a single \sersic\, function 
for the whole galaxy, (2) a bulge-disc decomposition model using a de 
Vaucouleurs ($R^{1/4}$) bulge plus exponential disc, (3) a bulge-disc 
decomposition model using a \sersic\, ($R^{1/n}$) bulge plus exponential disc. 
Repeat observations for $\sim700$ galaxies demonstrate that stable 
measurements can be obtained for object components with a half-light radius 
comparable to, or larger than, the seeing {{\it half}}-width at half maximum. 
We show that with careful quality control, robust measurements can be obtained 
for large samples such as the MGC. We use the catalogues to show that the 
galaxy colour bimodality is due to the two-component nature of galaxies (i.e. 
bulges and discs) and {{\it not}} to two distinct galaxy populations. We 
conclude that understanding
galaxy evolution demands the routine bulge-disc decomposition of the giant
galaxy population at all redshifts.
\end{abstract}


\section{Introduction}
\label{sec:intro}
 
To understand the origins of the diverse galaxy population observed today it 
is essential to quantify the properties of different 
structures which may be associated with separate formation processes. 
Recent work \citep[e.g.][]{baldry} reported that the local galaxy population
consists of two distinct classes (red and blue). 
\citet{driver_bimod} demonstrate that the galaxy bimodality is particularly
distinct in the colour-$\log(n)$ plane, where $n$ is the \sersic\, index for 
the best-fitting $R^{1/n}$ model to the 2D galaxy light distribution. 
Furthermore, they suggest that the bimodality may be better interpreted as a 
representation of the two-component nature of galaxies (i.e. red bulges and 
blue discs). They arrive at this conclusion by noting that E/S0's and Sd/Irr's 
lie in the two distinct peaks, whereas the Sabc's (i.e. bulge+disc systems) 
straddle the two peaks. The colour bimodality may reflect the end-result of 
two different processes associated with bulge and disc formation.

It has long been known that although galaxies 
can cover an expanse of quite different morphologies, they have many prevalent 
features in common, most notably spheroids (or bulges) and discs. The 
prominence of these features is often used to classify galaxies and place them 
into an evolutionary scenario \citep[e.g.][]{hubbletf,devauc59,vandenB}. Early 
classification of galaxy morphology was based on the visual inspection of 
images, and although this can be useful, significant differences between 
individual classifiers have been observed \citep[e.g.][]{lahav95,naim95a}, and 
less subjective techniques are clearly desirable. 

In order to provide a more quantitative assessment of galaxy morphology, a 
number of different methods have been proposed and employed. For example,
the concentration of the stellar distribution 
\citep[e.g][]{morgan58,morgan59,morgan62,fraser72,devauc77,graham_tc}, along 
with its asymmetry in a $C-A$ system \citep{burbidge,elmegreen,doi,schade,abraham96} can be used to quantify galaxy morphology in a way that is not dependent 
on an assumed model. Extended classification systems also include a measure of 
the `clumpiness' of the galaxy light \citep[$C-A-S$;][]{conselice03}. 
Other methods employ artificial neural networks 
\citep[e.g.][]{naim95b,lahav96,odewahn} to classify galaxies according to a 
Hubble sequence scheme.

Bulge-disc decomposition is another popular and useful method for quantifying 
the morphologies of galaxies by fitting model surface brightness profiles to 
data. Galaxies are described in terms of the two most easily recognised 
stellar components: bulges and discs. Such an approach has some clear 
disadvantages; the fitting process is model-dependent and does not account for 
secondary features such as spiral arms, rings, bars, star forming knots, tidal 
tails, and other asymmetries. 
However, even at $z\sim1$, bulges and discs are clearly in place and appear to 
be the dominant structural features in the majority of luminous ($M_{B}<-17$
mag) galaxies \citep[e.g.][]{simard99,ravindranath,barden,koo05,trujillo05}. 
In addition, software is now available which makes it fairly straightforward 
to perform bulge-disc decomposition on many thousands of galaxies 
\citep[e.g.][]{simard,peng,trujillo_ag,budda}.

A bulge-disc model is also convenient because the surface brightness profiles 
of bulges are known to be well modelled  by $R^{1/4}$ \citep{devauc}, 
$R^{1/n}$, \sersic\, \citep{sersic}, or exponential laws 
\citep{andred94,dejong96} and discs are observed to follow a pure exponential 
profile \citep{devauc59,freeman}. Bulges and discs are also often observed to 
have different average colours, metallicities, and kinematics, justifying their
treatment as distinct entities. Most models of galaxy formation and evolution 
also involve separate formation scenarios for bulges and discs 
\citep[e.g.][]{cole2000}. 

Using HST imaging, several statistically significant samples of high redshift 
galaxies with measured structural parameters have been constructed
\citep[e.g.][]{simard,ravindranath,barden}. However, for the local Universe
there only exist samples of hundreds of galaxies, that have had a bulge-disc
decomposition, and these are often preselected to be only late-type 
\citep[e.g.][]{dejong96,graham_deblok,macarthur}, or early-type 
\citep[e.g.][]{caon93,graham_guzman,dejong04}, or in high-density environments 
\citep[e.g.][]{gutierrez,christlein05}. In order to obtain a representative 
low redshift sample it is necessary to draw on a survey that is both deep and 
wide. \citet{blanton2003} fit a single \sersic\, function to over
180,000 SDSS galaxies, and recently \citet{tasca} have performed 
bulge-disc decomposition on a smaller sample of 1588 SDSS galaxies. 

In this paper we use a publically available bulge-disc decomposition code
\citep[GIM2D;][]{simard} to provide a quantitative measure of the surface 
brightness profiles of 10095 galaxies with $\bmgc<20$ mag in the Millennium 
Galaxy Catalogue (MGC). In Section \ref{sec:mgc} we briefly describe the MGC 
dataset analysed in this paper. Section \ref{sec:decomp} outlines the GIM2D 
fitting process. The resulting structural catalogues are introduced in
Section \ref{sec:cats}, and in Sections \ref{sec:quality}, 
\ref{sec:interpret}, and \ref{sec:repeat}, we address the validity and 
repeatability of the structural measurements. Finally, in Section
\ref{sec:bulgedisc} we recover the fundamental empirical results describing
bulges and discs (the Kormendy relation and the $\mu_{0}-\log(h)$ relation), and show 
that galaxy bimodality can be explained as a manifestation of distinct bulge and 
disc properties. In future papers we use the structural catalogues presented in
this paper to consider the bulge and disc luminosity functions (Allen \etal\, 
in prep 2006), the bulge-disc bivariate brightness and size distribution (Liske
\etal\, in prep 2006), and the supermassive black hole mass function (Graham 
\etal\, in prep 2006). Throughout this paper we assume a cosmology with 
$\Omega_{m}=0.3$, $\Omega_{\Lambda}=0.7$, and adopt 
$h=H_{0}/$(100 km s$^{-1}$ Mpc$^{-1}$) for ease of comparison with other 
results.

\section{The Millennium Galaxy Catalogue}
\label{sec:mgc}

The Millennium Galaxy Catalogue \citep[MGC\footnote{http$://$www.eso.org$/\sim$jliske$/$mgc};][]{liske} is a deep 
($\mu_{lim}$=26 B mag arcsec$^{-2}$), wide area ($\sim$37.5 deg$^{2}$) 
imaging and redshift survey covering a 0.5 deg wide strip along the equatorial 
sky from 10h to 14h 50$^{\prime}$. The survey region overlaps both the 
Two-degree Field Galaxy Redshift Survey \citep[2dFGRS;][]{colless}, and the 
Sloan Digital Sky Survey data release 1 \citep[SDSS-DR1;][]{abaz}. Comparison 
between these surveys shows that the MGC (which contains 10095 galaxies with 
$B<20$ mag) is deeper, more complete, more precise, and of higher resolution 
than either the 2dFGRS or SDSS-DR1 data sets \citep{cross,driver05}. In 
\citet{driver05}, their Figure 1 compares MGC imaging to DSS and SDSS imaging, 
and their Figure 3 demonstrates the extremely high redshift completeness 
of the MGC as a function of apparent magnitude, effective surface brightness, 
and colour.

Although the input MGC imaging is in the $B$-band, 
a full match to the SDSS photometric catalogues provides additional colour 
information in the $ugriz$ bands. Full details of the MGC, including 
observations, data reduction, image detection and classification, are given 
in \citet{liske}. The MGC redshift survey is discussed in detail in 
\citet{driver05}. With photometric precision of $\pm$0.03 mag, astrometric
accuracy of $\pm$0.08 arcsec \citep{liske}, and 96\% redshift completeness to 
$B_{{{\rm MGC}}}=20$ mag (increasing to 99.8\% for $B_{{{\rm MGC}}}<19$ mag), 
the MGC represents an extremely high quality and high completeness census of 
the local galaxy population. It is therefore the ideal data set to use for a 
detailed analysis of the structural composition of galaxies in the local 
Universe, and to provide a low redshift anchor for higher redshift studies.

\begin{figure}
\begin{center}
{\leavevmode \epsfxsize=8.0cm \epsfysize=8.0cm \epsfbox{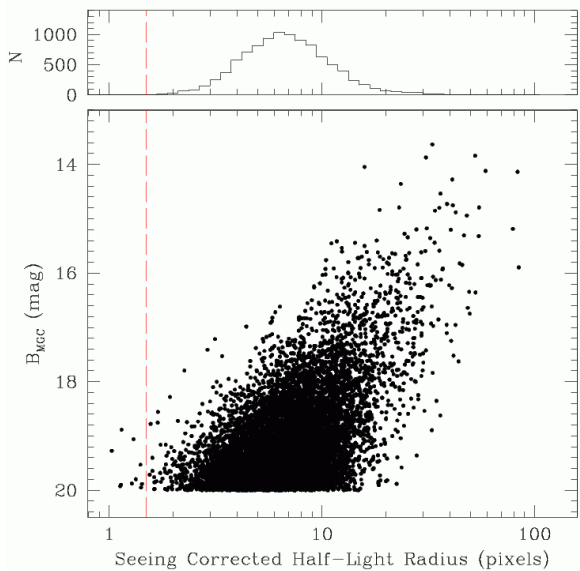}}
\end{center}
\caption{Seeing corrected half-light radii (pixels) for MGC galaxies 
\citep[as measured in][]{liske} versus apparent $B$-band magnitude. The dashed 
line shows the typical seeing HWHM. The pixel scale is 0.333$^{\prime\prime}$.}
\label{fig:mhlr}
\end{figure}

Figure \ref{fig:mhlr} shows the seeing corrected half-light radii 
\citep[as measured in][]{liske} for MGC galaxies as a function of apparent 
$B$-band magnitude. Most galaxies are intrinsically larger than the PSF HWHM 
and are therefore useful for structural analysis (n.b. those galaxies 
that are much smaller than the typical PSF size are eventually
removed from the final analysis - see Section \ref{sec:logic}). 

\section{Decomposition of Galaxy Profiles}
\label{sec:decomp}

To perform 2D bulge-disc decomposition on the MGC sample we elected to use the 
GIM2D package \citep{simard}. GIM2D allows galaxies to be modelled using a 
single-component model, or a two-component bulge plus disc model. Where a two 
component model is used, the components are required to have a common spatial 
centre, but their luminosities are independent, allowing the calculation of a 
bulge-to-total ($B/T$) luminosity ratio for each galaxy. The intensity profile
of the spheroidal (bulge) components, $I_{b}(R)$, can be described using a 
S\'{e}rsic function \citep{sersic,graham_driver}: 

\begin{equation}
I_{b}(R)=I_{e}\exp(-b_{n}[(R/R_{e})^{1/n}-1]),
\end{equation}

\noindent
where the effective radius, $R_{e}$, encloses half the total bulge luminosity,
and $I_{e}$ is the intensity at the effective radius. The S\'{e}rsic
parameter, $n$, determines the shape of the profile. A profile with
$n=4$ corresponds to the $R^{1/4}$ model \citep{devauc} that was traditionally 
used to describe bright elliptical galaxies. The term, $b_{n}$, is 
defined such that $\Gamma(2n)=2\gamma(2n,b_{n})$, where $\Gamma$ is the
complete gamma function and $\gamma$ is the incomplete gamma function,
ensuring that $R_{e}$ encloses half the light for all values of $n$. The term
$b_{n}$ can be approximated by $1.9992n-0.3271$ \citep{capaccioli}\footnote{For
$n<0.36$ this approximation begins to fail \citep{ciotti}, leading to 
uncertainties of order 0.1 mag arcsec$^{-2}$ in surface brightness. 
Unfortunately, although more accurate functions are available 
\citep[e.g.][]{macarthur}, this approximation is a fixed feature of the GIM2D 
code.}.

For discs, the intensity profile, $I_{d}(R)$, can be modelled using an 
exponential function:

\begin{equation}
I_{d}(R)=I_{0}\exp(-R/h),
\end{equation}

\noindent
where $I_{0}$ is the central intensity and $h$ is the scale length
(identical to a S\'{e}rsic profile with $n=1$). Although GIM2D permits the
application of a simple opacity model for discs, we choose at this stage to
treat discs as transparent and infinitely thin. We explore the effects of dust 
on disc opacity in a future paper (Allen \etal, in prep).

The ellipticity of the bulge component and the inclination of the disc 
component are permitted to vary independently. Flattening of bulges is 
described by an ellipticity, $\epsilon=1-b/a$, where $a$ and $b$ are the 
semi-major and semi-minor axes of the ellipse respectively. The bulge 
effective radius, and 
disc scale-lengths computed by GIM2D correspond to the semi-major axes of the 
bulge and disc respectively. Discs can have an inclination, $i$, from face-on 
($i=0$) to edge-on ($i=90$), measured under the assumption that face-on discs 
are circular. Bulges and discs are also permitted to have independent 
position angles, $\theta_{b}$ and $\theta_{d}$. Finally, the GIM2D code 
allows us to recalculate the background levels for each galaxy before the 
surface brightness models are fitted to the galaxies.

\subsection{Galaxy Input Images and Masks}
\label{sec:input}

\begin{figure*}
\begin{center}
{\leavevmode \epsfxsize=16.0cm \epsfysize=22.0cm \epsfbox{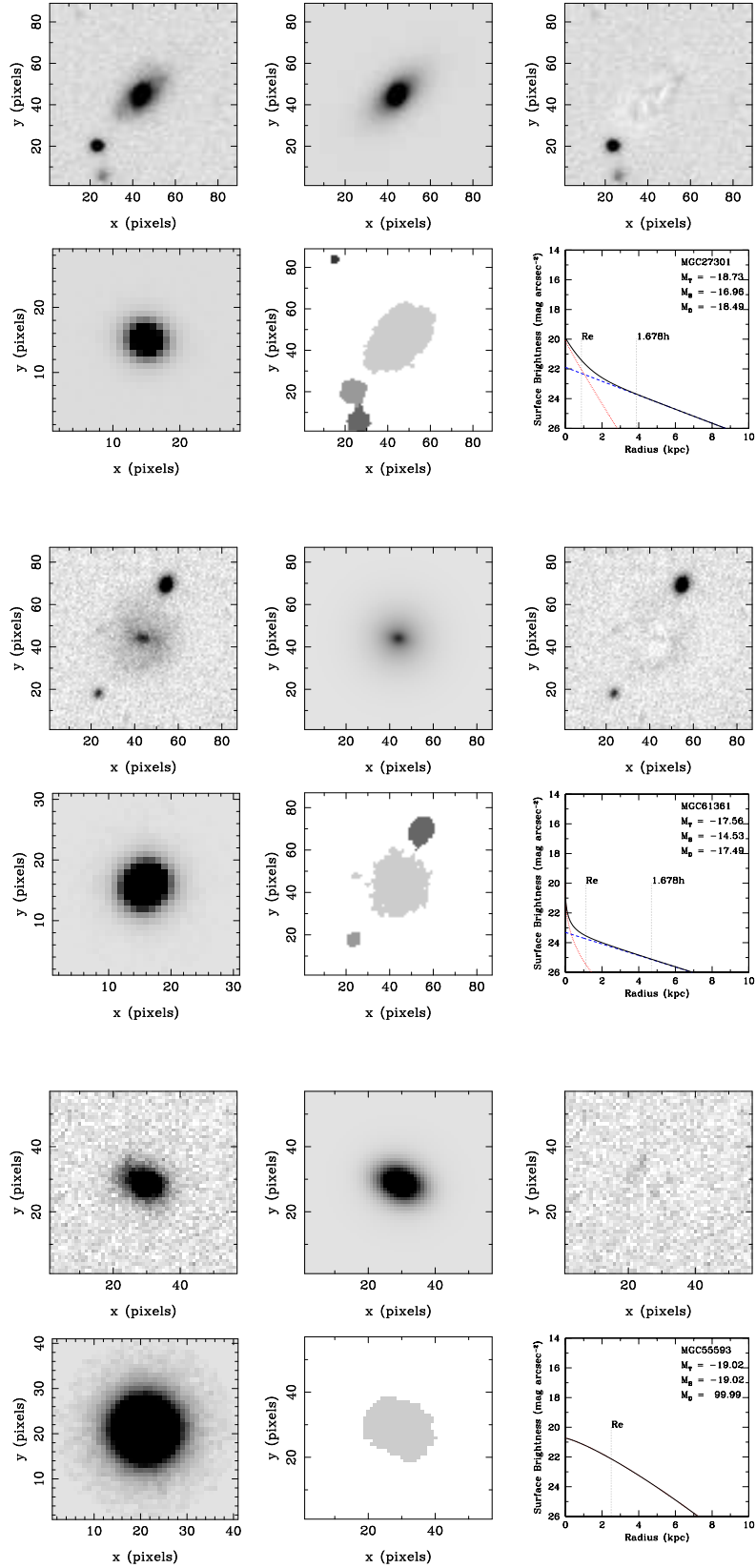}} 
\end{center}
\caption{GIM2D input and output for three representative example galaxies: MGC27301, MGC61361
(both shown with \sersic+exponential fits), and MGC55593 (single-component 
\sersic\, fit). For each galaxy the top row contains images of the 
galaxy (left), GIM2D model image (centre), and a residual image showing the
difference (right). The bottom row shows the input PSF (left, note the 
expanded scale), the SExtractor segmentation image or mask (centre), and a 
plot of the best-fitting profile found by GIM2D (right).}
\label{fig:outputs}
\end{figure*}

Initially, GIM2D uses the SExtractor \citep{bertin} output catalogue to prepare
galaxies for analysis. In particular, the galaxy $x$--$y$ position, sky 
background, and isophotal area as determined by SExtractor are required by 
GIM2D. Postage stamp images were created for each galaxy, centred on the input 
$x$--$y$ positions, with the height and width set to contain 10 times the 
SExtractor determined isophotal area of the galaxy. 

For each original image, SExtractor was set to produce a segmentation image, 
or mask. This consists of a pixel map defining background pixels to have a 
value of 0, and pixels that are considered part of objects are assigned values 
of 1 to $N$.
A mask postage stamp image with the same dimensions as the input image
was then produced for each galaxy. Figure \ref{fig:outputs} shows example
input MGC galaxies and their associated masks.
In some cases the masks did not distinguish between nearby 
objects (in most cases two or more nearby objects were given the same pixel 
value in the mask). As we discuss later, this results in an
erroneous output. When this occurred, SExtractor deblend 
and/or threshold parameters were changed to produce a correct mask (see 
Section \ref{sec:integrity} for full details) and the input masks remade.

\subsection{Point Spread Functions}

Before accurate profile measurements can be made, it is important to 
disentangle the intrinsic morphologies of galaxies from distortions that 
arise from the combined optical system of telescope, instrument, and 
atmosphere. These distortions vary both as a function of position and time, so 
their combined effects must be modelled using a point spread function (PSF), 
which is unique to each galaxy. In the fitting process, GIM2D convolves the 
model profiles with a model PSF before comparing the model to the data.

Initially the {\tt IRAF/DAOPHOT} package was used to create a PSF model for 
each CCD frame. This was achieved using a minimum of 20 stars (often more), 
selected on the basis of their SExtractor stellarity parameter being $>0.8$, 
and their magnitude lying in the range $17<B_{{{\rm MGC}}}<19.5$ mag. The stars
were chosen to provide good spatial coverage of the CCD, to not be blended, 
to have no bright neighbours or saturated pixels, and to be away from
CCD defects. The {\tt DAOPHOT} routine {\tt PSF} was then used to fit 
a model PSF for each CCD, based on the surface brightness profiles of the 
stars.

The PSF for each image was modelled using an analytic function, and a lookup 
table of residuals to account for the variation of the PSF across the CCD 
frame. The PENNY2 function, which consists of an elliptical Gaussian core with 
Lorentzian wings (both of which have independent position angles) was used to 
model the PSF, with the fitting radius of the function (i.e. the radius which 
encloses the region of pixels used in fitting the analytic function) defined 
to be equal to the median seeing for the frame. Where the median seeing is the 
median FWHM of simple Gaussian fits to the profiles of stars. The diameter of 
the region used for fitting the PENNY2 function is therefore twice the 
estimated FWHM. The PENNY2 function consists of five free parameters, $P_{n}$ 
(where $n=1...5$), and a normalisation factor, $A$, and can be expressed as:

\begin{equation}
{{\rm PENNY2}}=A\times\left(\frac{1-P_{3}}{1+z}+P_{3}\exp(-0.693e)\right),
\end{equation}

\noindent
where

\begin{equation}
z=\frac{x^{2}}{P_{1}^{2}}+\frac{y^{2}}{P_{2}^{2}}+xyP_{5},
\end{equation}

\noindent
and

\begin{equation}
e=\frac{x^{2}}{P_{1}^{2}}+\frac{y^{2}}{P_{2}^{2}}+xyP_{4}.
\end{equation}

\noindent
The resulting PSF image is the best-fitting function to the light profile of 
the stars in the frame weighted by their signal-to-noise.

To check the quality of the PSF, the {\tt ALLSTAR} routine was used to 
reconstruct and subtract the PSF from the stellar images. When the PSF is 
good, the stars are cleanly removed from images leaving only sky noise. In 
most cases the PSF provided a good model for the stars 
($\chi^{2}_{{{\rm red}}}\sim1$, where $\chi^{2}_{{{\rm red}}}$ is the reduced
$\chi^{2}$ for the fit), although for a small number of frames 
the $\chi^{2}_{{{\rm red}}}$ values were poor and additional stars were
rejected, and the PSF regenerated until an acceptable $\chi^{2}_{{{\rm red}}}$ 
was obtained. 

Using the model PSFs made for each CCD image, with the associated look-up
tables, the {\tt DAOPHOT} routine {\tt SEEPSF} was employed to create an image 
of an artificial star corresponding to the exact location of every MGC galaxy. 
GIM2D then convolves this image with the model fits before comparison 
with the input data. Example PSF stars are shown in Figure \ref{fig:outputs} 
(note the expanded scale for the PSFs).

\subsection{The GIM2D Models}

Three different models were fit to each MGC galaxy.

\begin{itemize}
\item One component: fitted with a S\'{e}rsic function.
\item Two components: de Vaucouleurs bulge + exponential disc.
\item Two components: S\'{e}rsic bulge + exponential disc. 
\end{itemize}

\noindent

A full description of the GIM2D fitting algorithm is given in \citet{simard},
but a brief outline follows. The first stage in the fit is to further refine 
the background subtraction. GIM2D uses those pixels flagged in the SExtractor 
segmentation image as `background' to estimate the mean value of the 
background. A 5 pixel buffer is placed around each object in the 
segmentation image to exclude possible object pixels from isophotes lower than 
the threshold that could potentially bias the background estimation. The new 
background value is then fixed, and 2D surface brightness fits are applied to 
the galaxies.

Table \ref{tab:ranges} shows the 8-12 parameters that are fit by GIM2D and the 
the upper and lower limits permitted for each of the fits, most of which are 
based on GIM2D defaults. Since some galaxies are known to have $n>4.0$ 
\citep[e.g.][]{caon93,graham96}, we permit the \sersic\, index 
to be as high as 12 \citep[c.f.][who fix $n_{{{\rm upper}}}=4$]{tasca}. 
Since we have assumed an infinitely thin disc, inclinations are given an upper 
limit of 85 degrees. Bulge ellipticities are permitted to have a maximum value 
of 0.7. However, in the \sersic-only catalogue the fit may best correspond to 
either a bulge or a disc. Therefore, in this catalogue, those objects with a 
GIM2D upper error limit of 0.7 for the ellipticity were refit using a revised 
upper limit of 0.92 (corresponding to an inclination of 85 degrees), and both 
versions of the fit stored. 

GIM2D uses the pixels flagged by the segmentation image as `object'
to measure image moments in order to produce initial estimates of the total 
flux, size, and position angle for each galaxy. Maximum values for these
parameters are then set at twice the best estimates from the image moments.
Using the limits applied in Table \ref{tab:ranges} and the maxima derived from 
the image moments, $N$ models are created coarsely sampling the permitted 
parameter space. For this analysis we set $N=400$. An `initial condition 
filter' is then used to search through these models to find the best fit. The 
values associated with this model are then used as the starting point for the 
final fit. GIM2D then uses the Metropolis algorithm to search for the maximum 
likelihood model (essentially a pixel-to-pixel $\chi^{2}$ minimisation) that 
best describes each galaxy. GIM2D also explores the parameter space around
the best fit to compute 68\% confidence intervals for each parameter.

\begin{table}
\begin{center}
\begin{tabular}{llcc} \\
\hline
Parameter & & Lower& Upper \\ \hline
Total Flux & $F$ (counts) & 0.0 & -- \\
Bulge/Total Flux Ratio & $B/T$ & 0.0 & 1.0 \\
Bulge Effective Radius & $R_{e}$ (pixels) & 0.0 & -- \\
\sersic\, Index & $n$ & 0.2 & 12.0\\
Bulge Ellipticity & $e$ & 0.0 & 0.7 \\
Bulge Position Angle & $\theta_{b}$ (degrees) & -360.0 & +360.0 \\
Disc Scale Length & $h$ (pixels) & 0.0 & -- \\
Disc Inclination & $i$ (degrees) & 0.0 & 85.0 \\
Disc Position Angle & $\theta_{d}$ (degrees) & -360.0 & +360.0 \\
$x$ --Position&$\Delta\,x$ (pixels) & -- & -- \\
$y$ --Position&$\Delta\,y$ (pixels) & -- & -- \\
Background level & $\Delta\,b$ (counts) & -- & -- \\
\hline\\
\end{tabular}
\caption{The 12 parameters that can be fitted by GIM2D, and the upper and lower
limits permitted in the fits. No entry indicates that no hard limits are 
applied. For the \sersic-only (one component) fit, the $B/T$ lower 
limit is constrained to 1.0, so only the \sersic\, component is fit. 
When a de Vaucouleurs ($R^{1/4}$) fit is applied, the upper and lower limits 
of the S\'{e}rsic index are held fixed at $n= 4.0$.}
\label{tab:ranges}
\end{center}
\end{table}

\begin{figure*}
\begin{center}
{\leavevmode \epsfxsize=15.0cm \epsfysize=5.0cm \epsfbox{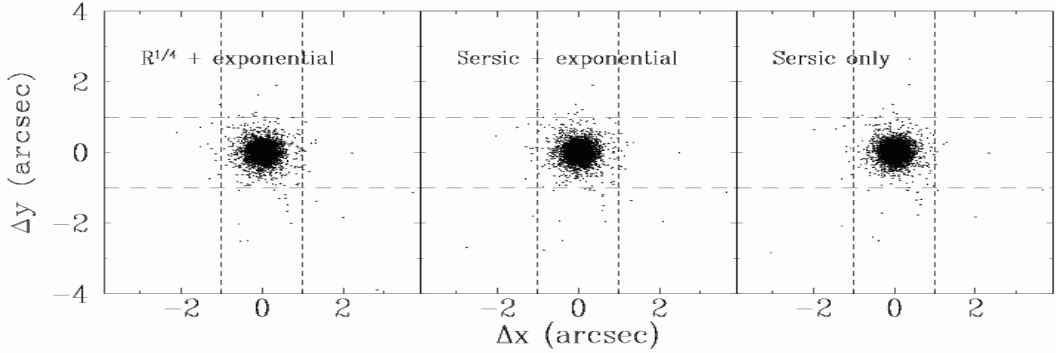}} 
\end{center}
\caption{Residual offsets in the location of the object centre as determined
by SExtractor and GIM2D (in $x$-$y$ pixels), are shown for the de Vaucouleurs + 
exponential catalogue (left), the \sersic\, + exponential catalogue (centre), 
and the single-component \sersic\, catalogue (right). The dashed lines 
correspond to a difference of 3 pixels (= 1 arcsec) in either $x$ or $y$.}
\label{fig:xyc}
\end{figure*}

\subsection{MGC Structural Catalogues}
\label{sec:cats}

Publically available MGC catalogues have already been released containing
non-GIM2D parameters, such as photometry, imaging, estimates of galaxy 
half-light radii, mean effective surface brightnesses, and a match to SDSS-DR1 
\citep{liske}. These were supplemented with redshifts and rest-frame colours in 
\citet{driver05}. The best-fitting values for the twelve parameters computed 
by GIM2D (see Table \ref{tab:ranges}) have been combined with the MGC-BRIGHT 
catalogue (i.e. 10095 galaxies with $B<20$ mag) to produce three structural 
catalogues: one for each of the three models fitted here. For each fit, GIM2D 
also measures the reduced $\chi^{2}$ (as computed between the PSF-convolved 
best-fitting model and the input data) as a measure of goodness-of-fit. 
Furthermore, using the topology of parameter space explored during the fitting 
process, GIM2D computes 68\% confidence limits for each of the parameters 
\citep[see][]{simard}. Although these errors are included in the catalogues, 
they only account for uncertainties in the formal fits. Uncertainties and
errors due to the sky background and PSF, which can often dominate, are not
considered in these estimates. This is discussed further in Section 
\ref{sec:repeat}. 

For each galaxy, in addition to the catalogue of structural parameters, output
PSF convolved model images, and residual images are also produced by GIM2D.
Figure \ref{fig:outputs} shows example output and residual images for three
galaxies: MGC27301, MGC61361, and MGC55593, along with the raw images, masks
and PSFs, and a plot of the best fitting profiles. Further details of our final
catalogue parameters are presented in Appendix A\footnote{Structural 
catalogues and images are made available at the MGC website
http$://$www.eso.org$/\sim$jliske$/$mgc}.

\subsubsection{Other Parameters}

After profile fitting has taken place and output and residual images made, 
GIM2D also computes several measures of concentration and asymmetry which are
described in detail in \citet{simard}. These include the $C-A$ system 
\citep[based on][]{abraham96} and, using residual images, $R_{T}$ and 
$R_{A}$ \citep{schade}. Two other measures, $A_{z}$, and $D_{z}$, are also 
defined by \citet{simard} as part of the GIM2D package. Although these 
parameters are not discussed further in this paper, the $C-A$ parameters are 
made available in the final MGC structural catalogues, and the other asymmetry
measures are available on request.

\section{Quality Control}
\label{sec:quality}

\subsection{Comparison with Independent Measurements}
\label{sec:integrity}

As an initial test on the accuracy of the GIM2D output, simple global 
observables such as magnitudes, half-light radii, and $x-y$ centroid positions 
can be compared to their values measured using other means in the 
\citet{liske} MGC catalogues. In most cases the agreement is excellent, 
although there are a number of outliers, which can be a useful part of the 
quality control process as we discuss in Section \ref{sec:reanal}. 

Figure \ref{fig:xyc} shows the residuals between the 
SExtractor and GIM2D centroids. The majority of objects (88.0\%) lie within 
1 pixel of the SExtractor central position, and 99.1\% have GIM2D central 
positions within 1 arcsec (3 pixels) of the SExtractor position for the 
\sersic+exponential catalogues, and similar accuracy is found for the other 
two models used in this paper. In most cases an outlying object is simply a 
sign of an irregular galaxy with an offset or ill-defined core, although it 
can, as we discuss in the next section, indicate problems with the masks 
(SExtractor segmentation images), or the GIM2D output.

\begin{figure}
\begin{center}
{\leavevmode \epsfxsize=8.0cm \epsfysize=8.0cm \epsfbox{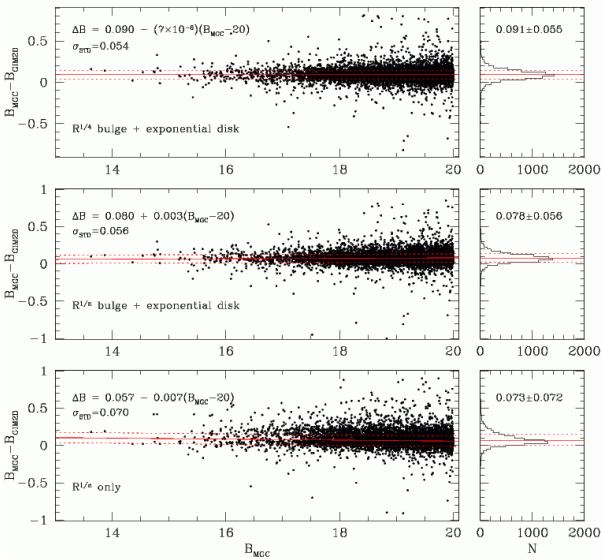}} 
\end{center}
\caption{Differences between MGC galaxy magnitudes (determined by SExtractor) 
and GIM2D magnitudes, as a function of MGC magnitude. The solid line shows the
least squares fit to the data, and the dashed lines define the $1-\sigma$
envelope. There is no apparent trend with magnitude.
The GIM2D magnitudes are slightly brighter in all three catalogues: 
de Vaucouleurs+exponential (top), \sersic+exponential (middle), and \sersic\, 
(bottom). The panels on the right show histograms of the residuals. The mean 
and standard deviation (with $3\sigma$ clipping) of these distributions are 
shown in each case.}
\label{fig:magc}
\end{figure}

Figure \ref{fig:magc} shows the difference between the \citet{liske} and 
GIM2D magnitudes as a function of MGC magnitude, and the best-fitting 
relations between these parameters along with residual
histograms, and the mean and standard deviation (with iterative $3\sigma$ 
clipping). For all three structural catalogues the GIM2D magnitudes are 
brighter by $0.07-0.09$ mag, with a $0.05-0.07$ mag standard deviation. There 
appears to be no significant trend with magnitude. 
Systematically brighter GIM2D magnitudes are not unexpected since profiles are 
integrated to $R=\infty$, and it is known that flux can be missed when using 
aperture or Kron magnitudes \citep[see][]{graham_driver}. There are a small 
percentage of objects with large residuals that we discuss in the next Section.

\begin{figure*}
\begin{center}
{\leavevmode \epsfxsize=15.0cm \epsfysize=5.0cm \epsfbox{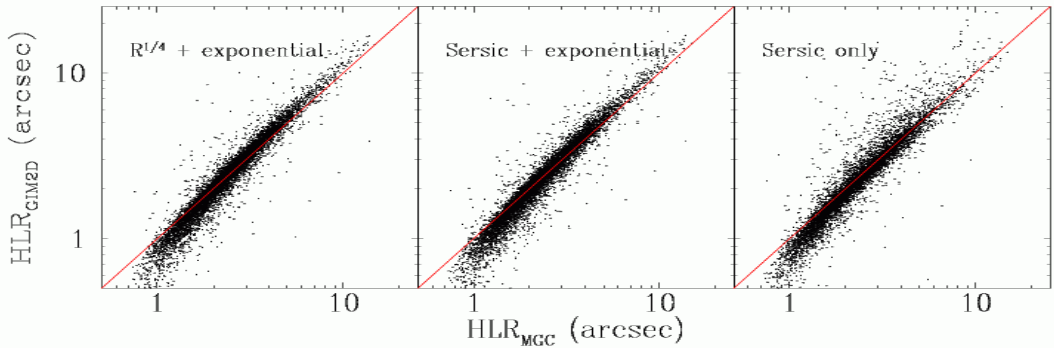}} 
\end{center}
\caption{Comparison between galaxy half-light radii measured by GIM2D, and 
seeing corrected half-light radii computed in \citet{liske}, for
the de Vaucouleurs+exponential catalogue (left), the \sersic+exponential 
catalogue (centre), and the single component \sersic\, catalogue (right). 
There is good agreement except at small radii where the PSF dominates.}
\label{fig:hlrc}
\end{figure*}

Finally, GIM2D also computes an {{\it approximate}} measure of the total 
galaxy half-light radius, which assumes that the bulge and disc have the same 
position angle (although the position angles are independent in the fits).
In Figure \ref{fig:hlrc} the GIM2D galaxy half-light radii are compared to the
seeing corrected half-light radii measured in \citet{liske}. Again, there is 
good agreement but with some scatter. It is also notable that at small radii 
($\lsim1.2''$) the GIM2D radii are smaller than the \citet{liske} radii. This 
is unsurprising as it is in the regime where the PSF is of comparable angular 
size to the galaxies. GIM2D measurements will be dominated by errors due to 
the PSF correction and the original MGC measurements will be subject to the 
assumptions of the seeing correction applied by \citet{liske}. For larger, 
well resolved galaxies the GIM2D half-light radii are slightly larger, as 
expected, given their brighter magnitudes.

\subsection{Reanalysis of Outliers}
\label{sec:reanal}

Analysis of the input and GIM2D output images of outlying objects in Figures 
\ref{fig:xyc} and \ref{fig:magc} reveals that the difference in parameters 
between \citet{liske} and GIM2D measurements can have a number of sources. In 
some cases galaxies are irregular, with more than one possible `centre', or 
have a very asymmetric surface brightness distribution. The GIM2D models can 
be a poor fit to such objects, and position and magnitudes consequently have
significant differences. In addition, many objects have nearby neighbours,
in which case SExtractor has applied a deblending algorithm to attempt to 
separate the flux from the two objects. SExtractor creates a mask 
image with overlap pixels simply designated to one object or the other. 
However, when SExtractor computes the magnitudes for these objects, the flux in
the pixels lying in overlap regions is shared between the two objects.
No such deblending is performed by GIM2D, which simply uses the mask, and all 
the flux is assumed to belong to one object or the other. 
Therefore some GIM2D inputs can be contaminated by light from nearby 
neighbours, and the derived magnitudes will differ from those determined by 
SExtractor.

Another cause of large parameter differences are erroneous masks or 
segmentation images. SExtractor, like most automated detection algorithms, is 
typically set to run with `optimal' tuning (i.e. the detection parameters 
are set so that reasonable output is produced for as many objects as 
possible). Nevertheless, a fraction of objects at the extremes will still be 
incorrectly recovered. Perhaps the best examples are the SExtractor deblending 
parameters {{\tt DEBELND\_NTHRESH}} and {{\tt DEBELND\_MINCONT}} which govern
the number of deblending thresholds, and the minimum contrast used when 
SExtractor decides whether two nearby local maxima in surface brightness 
are significant enough to be considered separate objects, or just structure 
within a single object. At one extreme, if the deblending is too strong, 
objects such as spiral discs or irregular galaxies can be broken up into 
several smaller objects. At the other extreme, two nearby, but clearly 
distinct objects are considered one, and assigned the same arbitrary pixel 
value in the mask (this is especially common when a small
object lies close to a much larger one).  To confuse matters further, in a 
small number of cases, SExtractor has failed to assign different pixel values 
to two nearby objects in the mask, even if two objects appear 
correctly deblended in the catalogue. These errors can be fixed by changing
the deblend parameters to be more or less sensitive depending on the problem,
and rerunning SExtractor. However it is not possible, even using the most 
optimal setup, to produce output that is satisfactory in every case. This is
a problem when trying to automate measurements for increasingly large
data sets as it becomes more and more difficult to check all objects by eye
(and defeats the purpose of trying to automate the process!).
Either some fraction of wrong detections has to be tolerated, or a way needs
to be found where the failures can be flagged and corrected.

For the MGC, all 10095 objects were checked by eye in \citet{liske}, and 
therefore, as already discussed, differences between the \citet{liske} 
catalogues and GIM2D output can be indicative of bad masks. If a 
galaxy has been over-deblended, then the GIM2D magnitude will be an 
under-estimate. When two galaxies are treated as one, the centroids will be 
grossly different, and the GIM2D magnitudes over-estimated. All objects with 
$\Delta x$ or $\Delta y$ greater than 3 pixels (1 arcsec) or a
magnitude difference $|\Delta B|>0.2$ mag, were selected and
inputs, outputs, and masks examined by eye. It was found that 222 
masks needed to be corrected by hand.

\begin{figure*}
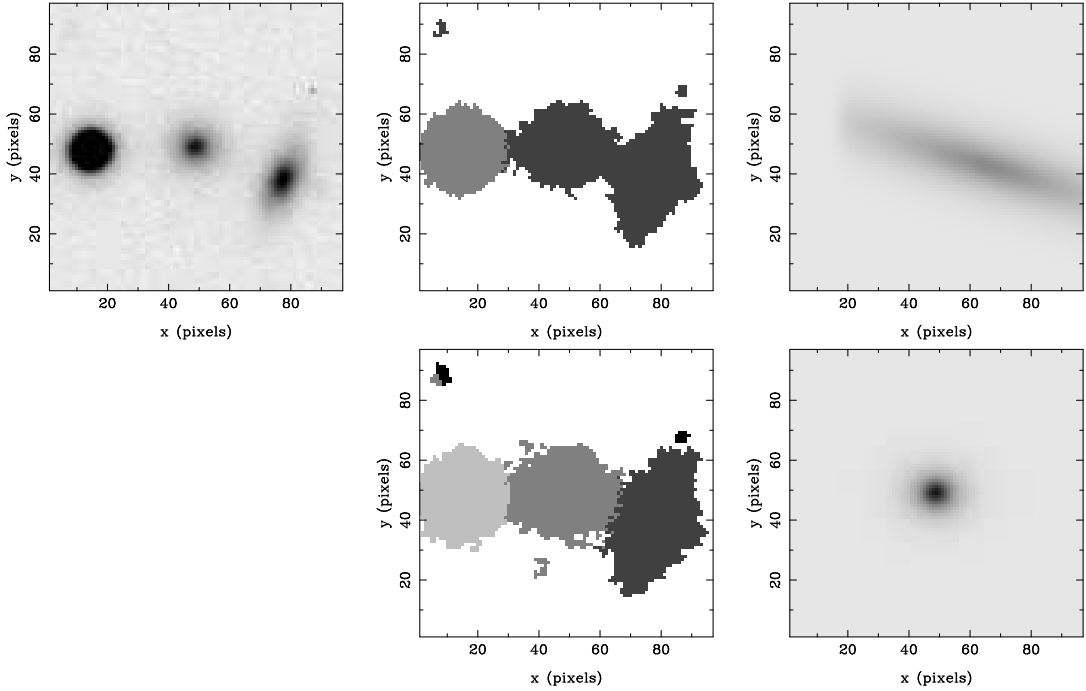

\begin{center}
\begin{tabular}{ccc}
{\leavevmode \epsfxsize=4.5cm \epsfysize=4.5cm \epsfbox[44 28 509 502]{badsegin.ps}} &
{\leavevmode \epsfxsize=4.5cm \epsfysize=4.5cm \epsfbox[44 28 509 502]{badsegm.ps}} &
{\leavevmode \epsfxsize=4.5cm \epsfysize=4.5cm \epsfbox[44 28 509 502]{badsego.ps}}\\
&
{\leavevmode \epsfxsize=4.5cm \epsfysize=4.5cm \epsfbox[44 28 509 502]{badsegmc.ps}} &
{\leavevmode \epsfxsize=4.5cm \epsfysize=4.5cm \epsfbox[44 28 509 502]{badsegoc.ps}}\\
\end{tabular}
\end{center}
\caption{An example of bad and corrected mask images. The top left
panel shows the input image for MGC64481 (the central object), and the
original mask (top centre). The top right image is 
the original GIM2D output image, which can be initially identified due to its 
large centroid offset, and magnitude difference between the GIM2D output and 
\citet{liske}. The bottom panels show the corrected mask and output images.}
\label{fig:badseg}
\end{figure*}

An example is shown in the top row of Figure \ref{fig:badseg}, with the 
original input image, bad mask (which assigns the same pixel 
value to two distinct objects), and the clearly erroneous GIM2D fit on the 
right. A corrected mask, which is produced by  making small adjustments to 
SExtractor deblending and threshold parameters, is shown in the bottom panel, 
along with the (now corrected) GIM2D output.

\section{Interpreting GIM2D output}
\label{sec:interpret}

One of the problems with automatic bulge-disc decomposition routines is that
in addition to perfectly good fits, they automatically generate a lot of 
rubbish. In this Section we describe our best efforts to clean up this 
situation.

As a starting point we first trim the catalogue and examine only those 
systems within the redshift range $0.013 < z < 0.18$, and impose the galaxy 
size and surface brightness limits introduced in \citet{driver05}. This results
in a trimmed catalogue of 7750 galaxies \citep[see][]{driver05}. 

The interpretation of single-component \sersic\, fits is fairly trivial, with 
the resulting light profile assumed to be the best-fit to the light from the 
entire galaxy. However, when two components are modelled, the two functions 
are typically taken to correspond to a bulge and a disc. It is 
necessary to verify whether or not this is the case, ensuring that the 
two-component model really does correspond to two distinct structural 
components within a galaxy. This is especially important in the case where a 
\sersic+exponential model is used because this fit has the most degrees of 
freedom.

\begin{figure*}
\begin{center}
\begin{tabular}{ccc}
{\leavevmode \epsfxsize=5.5cm \epsfysize=5.5cm \epsfbox{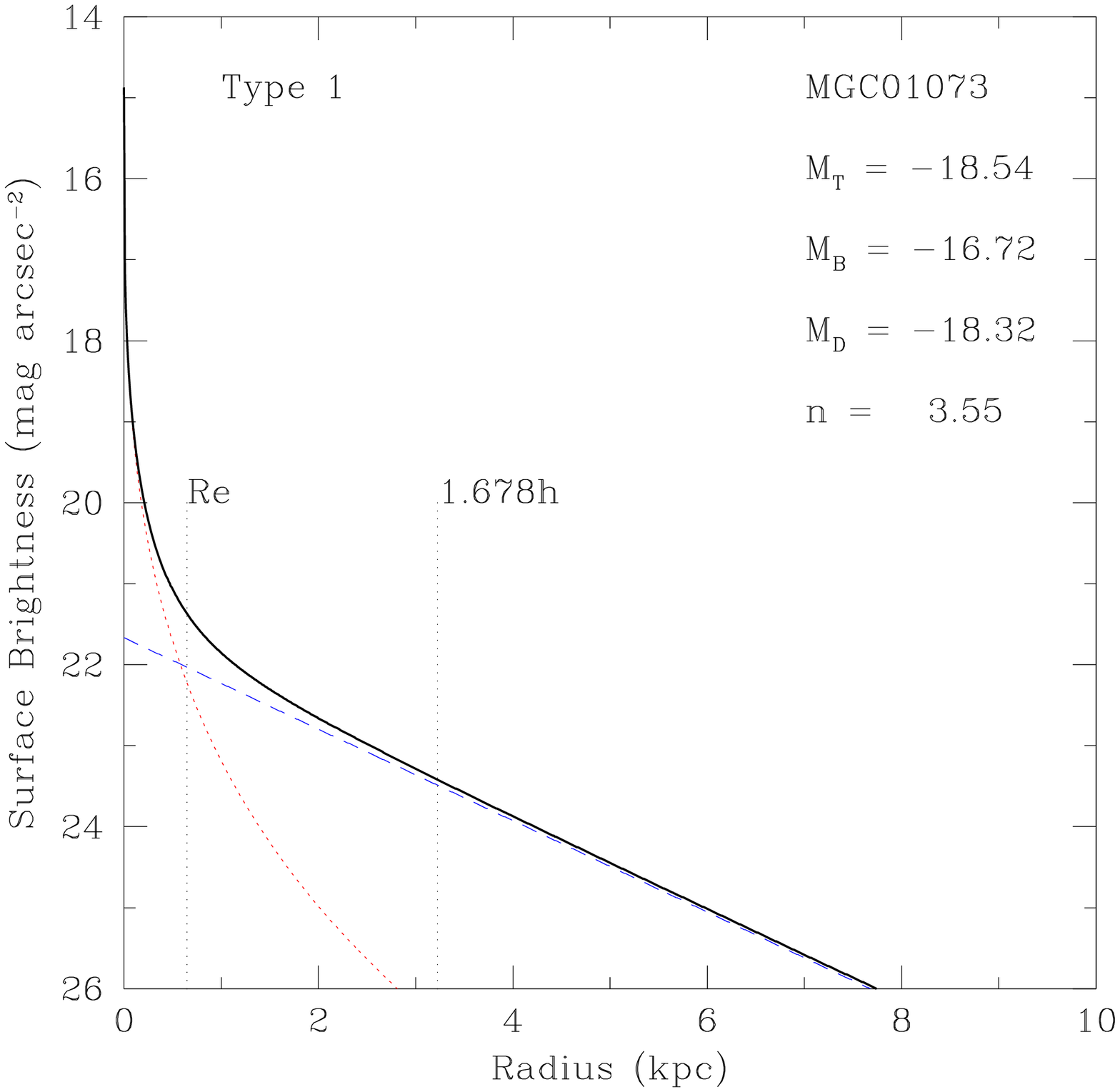}} &
{\leavevmode \epsfxsize=5.5cm \epsfysize=5.5cm \epsfbox{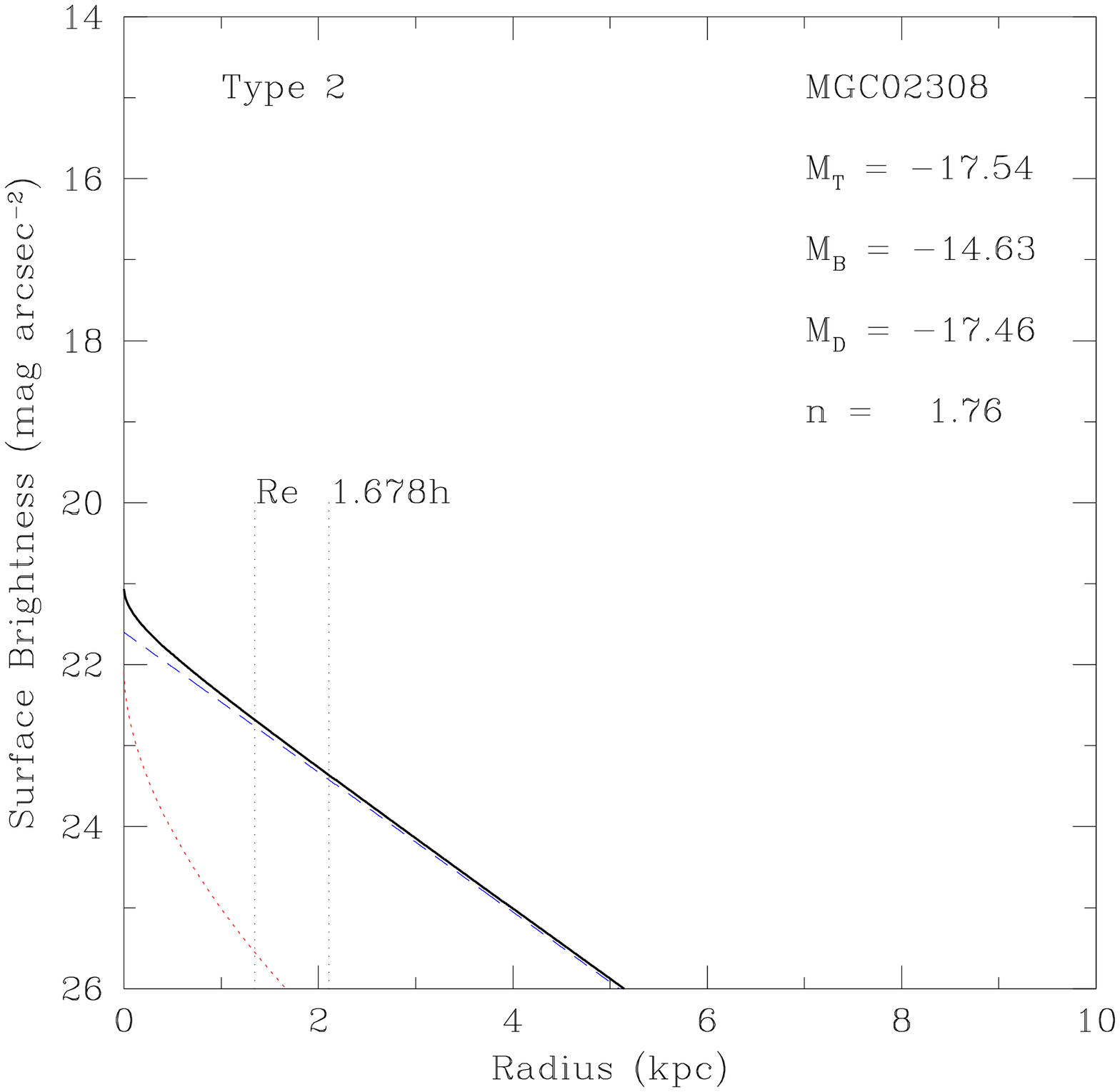}} &
{\leavevmode \epsfxsize=5.5cm \epsfysize=5.5cm \epsfbox{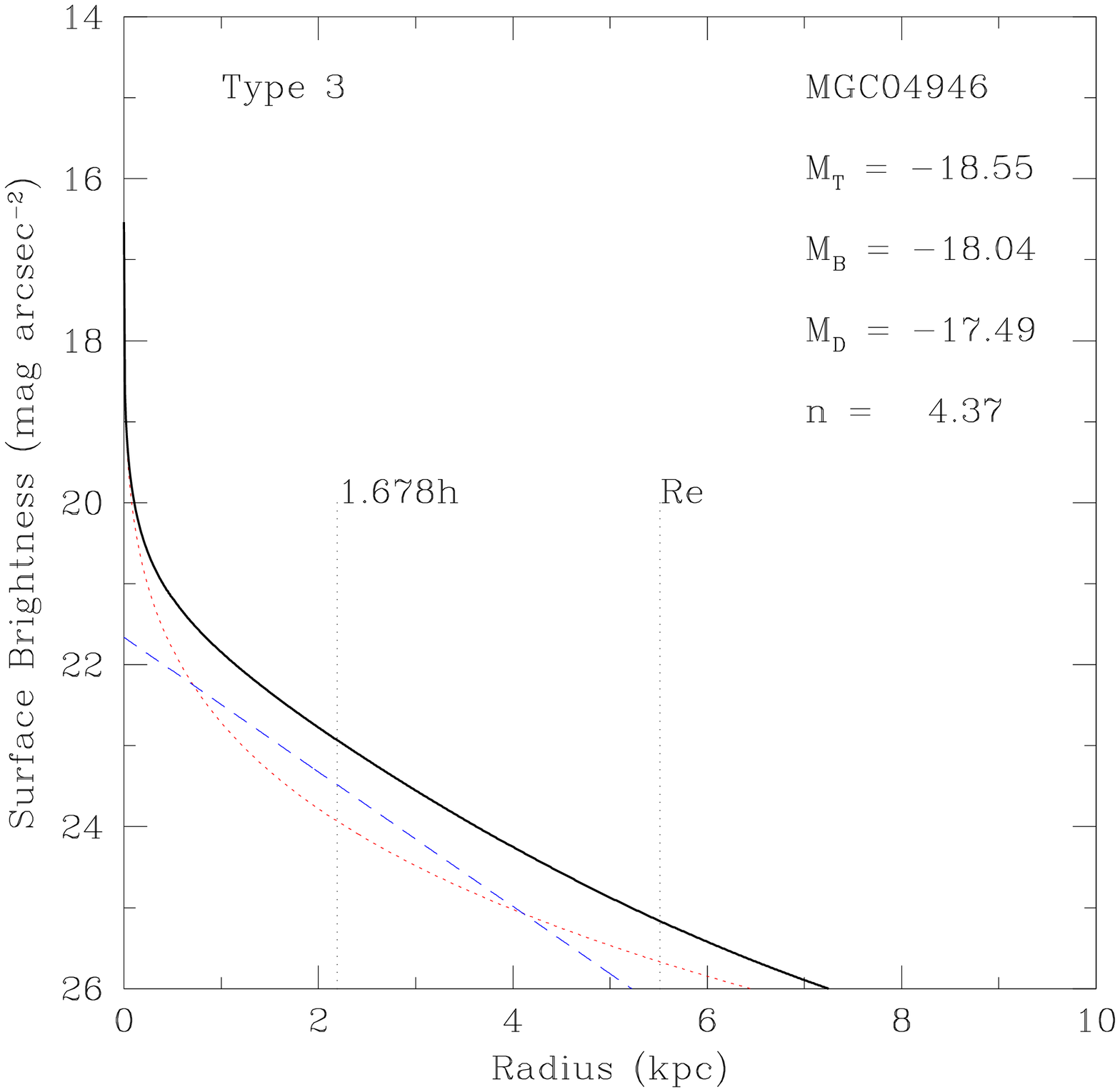}}\\
{\leavevmode \epsfxsize=5.5cm \epsfysize=5.5cm \epsfbox{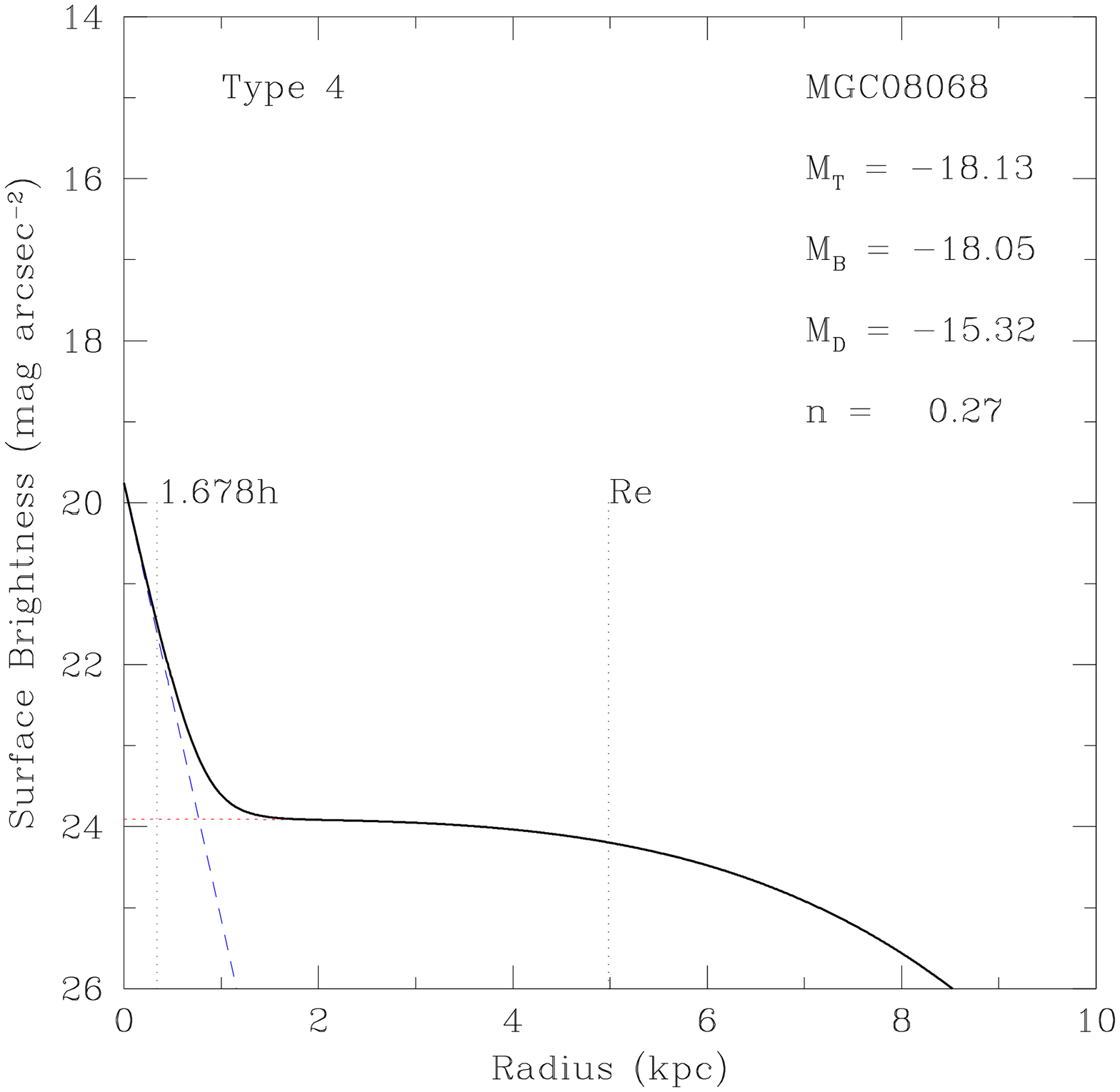}} &
{\leavevmode \epsfxsize=5.5cm \epsfysize=5.5cm \epsfbox{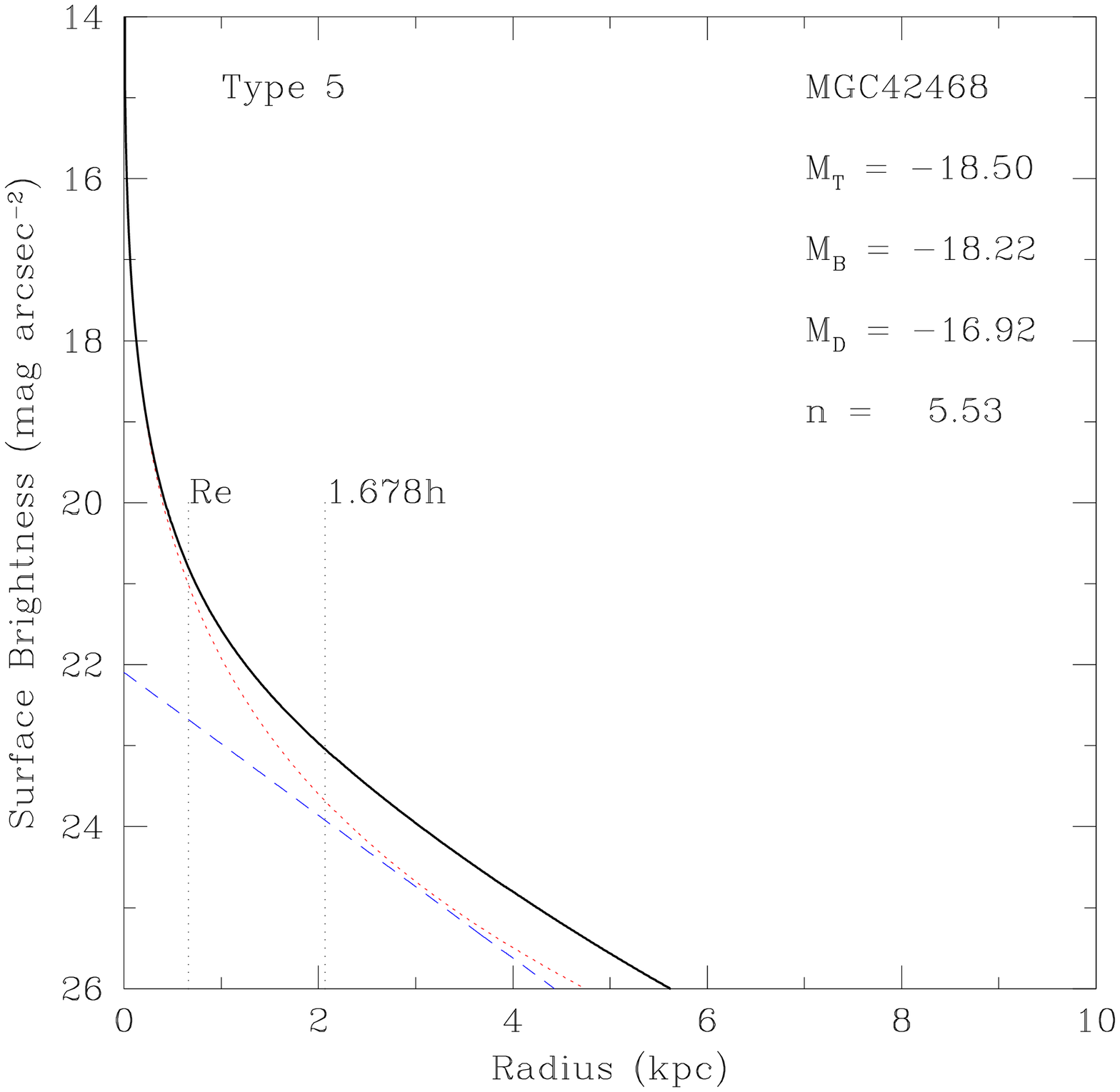}} &
{\leavevmode \epsfxsize=5.5cm \epsfysize=5.5cm \epsfbox{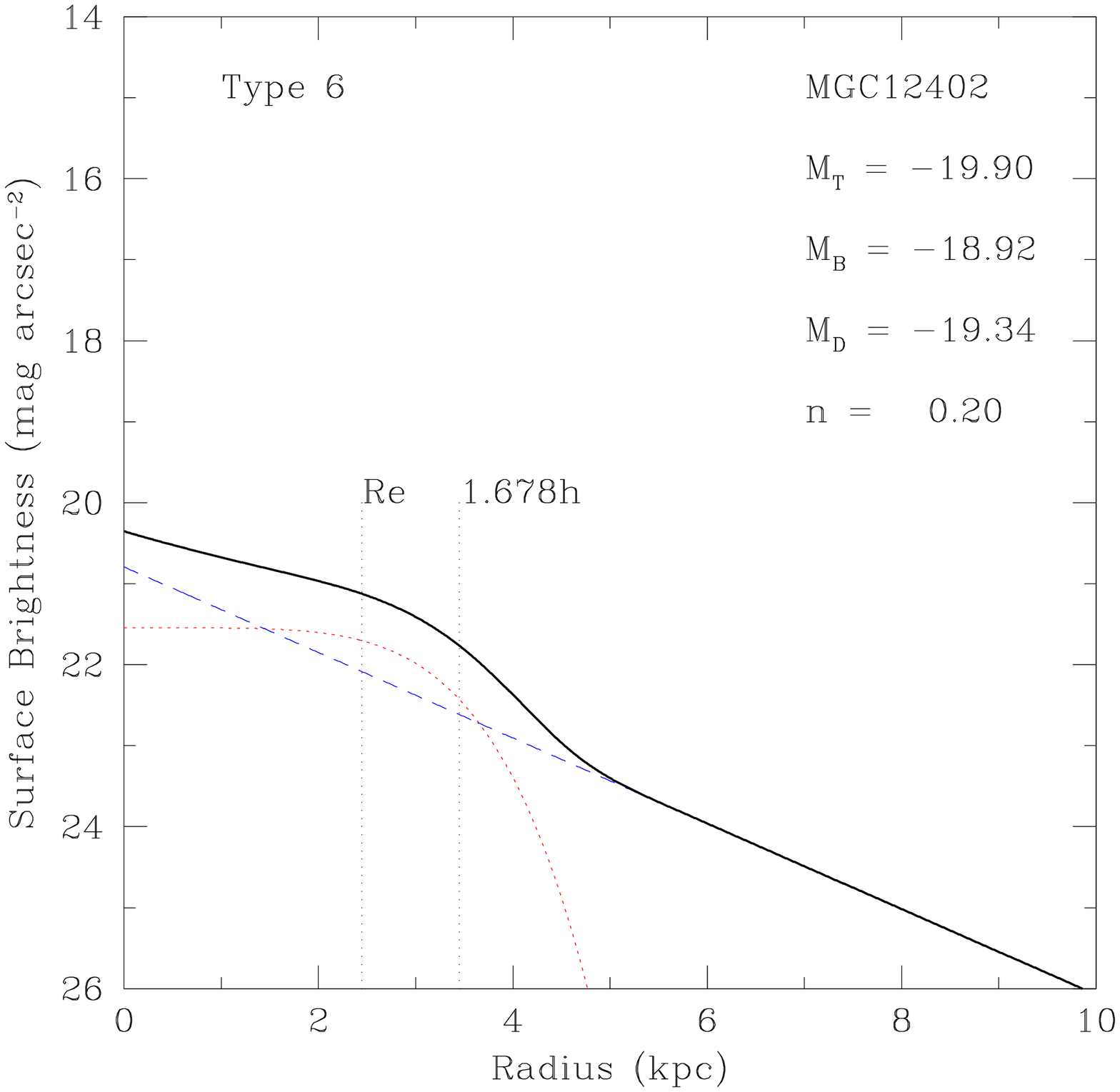}}\\
\end{tabular}
\end{center}
\caption{Examples surface brightness profiles for the 6 distinct types of 
GIM2D fit that we use to classify galaxies from the \sersic\, + exponential 
catalogue (see text). In each panel the bold solid line represents the total 
profile, the S\'{e}rsic component is represented by a dotted curve, and the 
dashed straight line represents the exponential component. The locations of 
the half-light radii for each component: $R_{e}$, and $1.678h$, are also 
shown. Table \ref{tab:types} summarises the frequency with which each type 
occurs.}
\label{fig:types}
\end{figure*}

No restrictions on the relative sizes of the `bulge' and `disc' components
were applied when using GIM2D. The two components can sometimes be 
inverted or used to fit other structural features, especially in irregular
galaxies. The component light profiles may also cross once, twice or not at 
all. To ensure the correct interpretation of the GIM2D fits (and whether
components should be interpreted as bulges, discs, or something else) we
classify all profiles into one of eight different types: six are composed of 
two components, and two have only single-component profiles. The two-component 
profile types are plotted in Figure \ref{fig:types} for six example galaxies 
from the \sersic\, + exponential catalogue. The profile types can be 
summarised as follows:

\begin{itemize}

\item {{\bf Type 1}} `Classic' profile. The S\'{e}rsic profile dominates at 
the centre, while the exponential profile dominates the flux at large radii. 
At a surface brightness brighter that 26 B mag arcsec$^{-2}$ the profiles 
cross only once (i.e. they have the same surface brightness at 
only one radius).



\item {{\bf Type 2}} Disc dominated system. The exponential profile dominates 
at all radii, with a small, central S\'{e}rsic profile. The profiles never 
cross, and $B/T$ is always $<$ 0.5. 

\item {{\bf Type 3}} S\'{e}rsic profile dominates at small and large radii, 
but an exponential profile dominates at intermediate radii. The profiles cross 
twice. $B/T$ is always $>0.5$, and the Sersic index, $n$ is typically $>1.5$.
Objects fitted by these profiles are typically single-component Elliptical 
galaxies.

\item {{\bf Type 4}} Inverted profile. The exponential profile dominates at 
the centre, and the \sersic\, profile dominates at large radii. The profiles 
cross only once, and $n$ is always small. The correct interpretation
of the profile is problematic, and could be a signature of possible
disc truncation, poor background subtraction, or the profile could correspond
to irregular or dwarf systems.

\item {{\bf Type 5}} Bulge or disc dominated. The \sersic\, profile dominates
at all radii, with a weak underlying exponential component. The profiles 
never cross, and $B/T>0.5$. There is a bimodal distribution in $n$ for this 
type (see Section \ref{sec:logic}). 

\item {{\bf Type 6}} Disc with perturbation. The exponential profile dominates 
at small and large radii, and the S\'{e}rsic profile briefly dominates at 
intermediate to large radii. The Profiles cross twice. The S\'{e}rsic 
component models either a disc perturbation (such as a spiral arm), or 
features in an Irregular galaxy. 

\item {{\bf Type 7}} Pure disc: an exponential-only profile with $B/T=0$. 

\item {{\bf Type 8}} Pure bulge, disc, or other: S\'{e}rsic profile only with 
$B/T=1$. A bimodal distribution in $n$ reveals discs with $n\sim1$, and bulges 
with $n\sim4$ for our sample.

\end{itemize}

All 8 profile types occur in the \sersic+exponential catalogue, types 
1,2,3,5,7 and 8 occur in the de Vaucouleurs+exponential catalogue, and the 
\sersic-only catalogue only contains type 8 profiles by definition. In both the
\sersic+exponential and the de Vaucouleurs+exponential catalogues the majority
of galaxies are of type 1. In the de Vaucouleurs+exponential catalogue 
$\sim1/3$ of galaxies are best modelled by an exponential-only fit (type 7),
three times as many as in the \sersic+exponential catalogue.
Table \ref{tab:types} summarises the fraction of each type that is found in 
the three different catalogues.

\begin{table}
\begin{center}
\begin{tabular}{cccc} \\
\hline
Type & $R^{1/n}+$Exp (\%) &  $R^{1/4}+$Exp (\%) & $R^{1/n}$ (\%) \\ \hline
1 & 50.8 & 52.5 & 0\\
2 & 9.5 & 0.4 & 0\\
3 & 1.9 & 4.1 & 0\\
4 & 5.9 & 0 & 0\\
5 & 14.1 & 7.1 & 0\\
6 & 4.8 & 0 & 0\\
7 & 12.1 & 35.3 & 0\\
8 & 0.8 & 0.5 & 100\\
\hline\\
\end{tabular}
\caption{The fraction of galaxies corresponding to each profile type (see 
Figure \ref{fig:types}).}
\label{tab:types}
\end{center}
\end{table}

\subsection{The Logical Filter}
\label{sec:logic}

The use of raw GIM2D profiles from the two-component models would clearly be 
inappropriate given the distinct and often physically meaningless profiles 
that are fitted in some cases (e.g., types 4 or 6). Moreover, many systems
do not actually exhibit two components, and in these cases it is not surprising
that a two-component fit produces erroneous output. A non-negligible fraction 
of incorrect output is typical for most automated algorithms. A quick 
inspection of the bulge-disc decompositions from BUDDA \citep{budda}, and
\citet{aguerri} reveal that problems with automatic decompositions are 
pandemic, with many fits having unrealistic bulge-to-disc size ratios,
disc components clearly falling faint of the outer discs in spiral galaxies,
and bulge flux dominating at large radii in late-type spiral galaxies, etc.
While automatic codes find the optimal mathematical solution,
perturbations in the real data from the fitted models often result
in the mathematical solution being an unphysical one.

However, it is possible to ask whether one can identify a suitable set of 
`rules' to apply to the GIM2D catalogues to produce a meaningful, final 
`filtered' catalogue. This catalogue would consist of bulge-disc decompositions
where appropriate (using the \sersic+exponential fits), and single-component 
\sersic\, fits otherwise. The interpretation of the single-component \sersic\, 
fits (as a bulge or a disc) would depend on colour and/or \sersic\, index.
This same approach could be followed using the de Vaucouleurs catalogue
for bulge-disc decomposition but we choose to consider only the more 
meaningful \sersic+exponential catalogues here.

For each fit GIM2D computes the reduced-$\chi^{2}$ as a measure of the
goodness-of-fit for the output model.
In most cases both the \sersic+exponential fits and the single-component 
\sersic\, catalogues have reasonable reduced-$\chi^{2}$ values ($\sim1$), 
although the \sersic+exponential reduced-$\chi^{2}$ values are generally 
smaller. Although the \sersic+exponential fits may have a reasonable
reduced-$\chi^{2}$, this doesn't indicate whether or not the fit is 
{{\it appropriate}} (i.e. whether the \sersic\, function is really 
corresponding to a bulge and the exponential function to a disc).
Therefore the approach followed here involves analysis of the global 
properties of each type (including colours, \sersic\, indices, and the sizes 
of components), along with visual examination of examples of each of the eight 
types. 

\begin{figure}
\begin{center}
{\leavevmode \epsfxsize=8.0cm \epsfysize=8.0cm \epsfbox{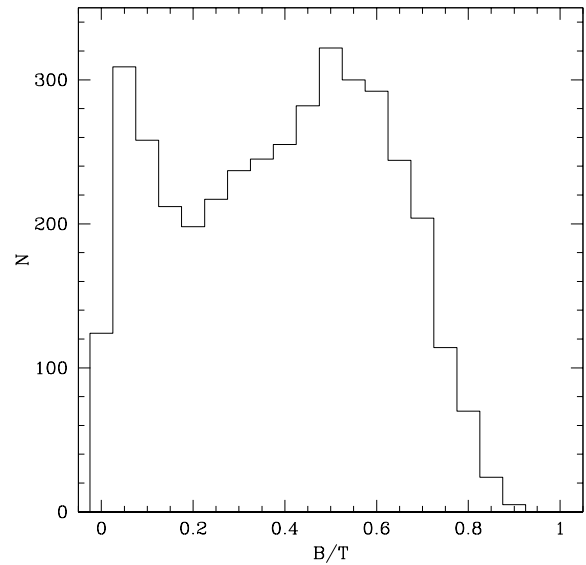}} 
\end{center}
\caption{The distribution in $B/T$ for (classic) type 1 galaxies. The lack
of objects with very high $B/T$ is notable.}
\label{fig:btdist}
\end{figure}

Objects have a range of $B/T$ values (Figure \ref{fig:btdist} shows the $B/T$ 
distribution for our type 1 galaxies), although we note that the frequency of 
high B/T values is small (indeed, only 62/7750 objects are type 8 with 
$B/T=1$). This perhaps reflects the fact that the addition of a weak 
exponential model to a light profile dominated by a \sersic\, model can make 
little difference to the overall total profile \citep[see e.g.][]{saglia97}. 
The exponential component may represent a weak disc, an isophotal twist, a PSF 
error, or some perturbation in the profile. This degeneracy means that 
kinematic measurements are required to test for the presence of a 
low-luminosity, rotationally supported disc.

\begin{figure}
\begin{center}
{\leavevmode \epsfxsize=8.0cm \epsfysize=8.0cm \epsfbox{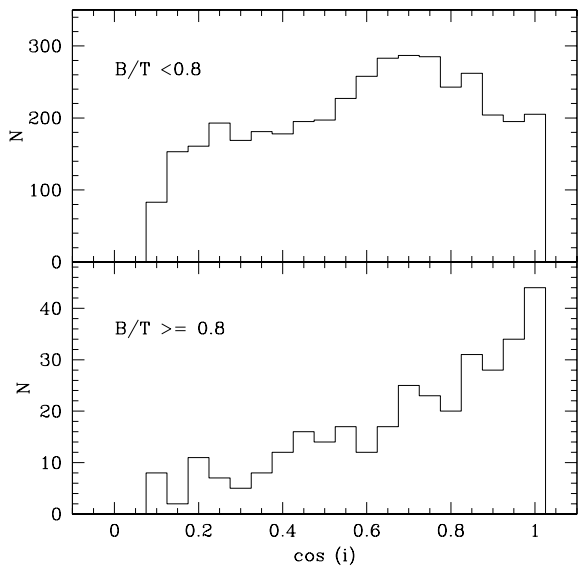}} 
\end{center}
\caption{Distribution in $\cos(i)$ for the `disc' component of galaxies with 
types 1 and 5 with a bulge component having $n>1.5$. The top panel shows a 
roughly flat distribution for those galaxies with $B/T<0.8$, whereas the lower 
panel, for galaxies with $B/T>0.8$, indicates a suspicious preference for 
face-on exponential profiles.}
\label{fig:inctest}
\end{figure}

Figure \ref{fig:inctest} shows the distribution of disc inclinations for 
spheroid dominated systems (type 1 and type 5 galaxies with $n>1.5$) for 
$B/T<0.8$ and $B/T>=0.8$. For objects with $B/T<0.8$, the distribution is 
relatively flat, as expected for a random sample of discs with random 
orientations in space. However, for $B/T>=0.8$, `discs' with inclinations 
towards face-on dominate, especially for those objects that have been visually 
classified as E/S0. It is therefore highly improbable 
these faint discs are all real. Therefore, we choose to replace the \sersic\, 
+ exponential fits for type 1 \& 5 objects with $B/T>0.8$ with single 
\sersic\, profiles \citep[c.f.][who replace objects with $B/T>0.6$ with 
single \sersic\, profiles]{trujillo_ag,gutierrez}.

\begin{figure*}
\begin{center}
{\leavevmode \epsfxsize=14.0cm \epsfysize=14.0cm \epsfbox{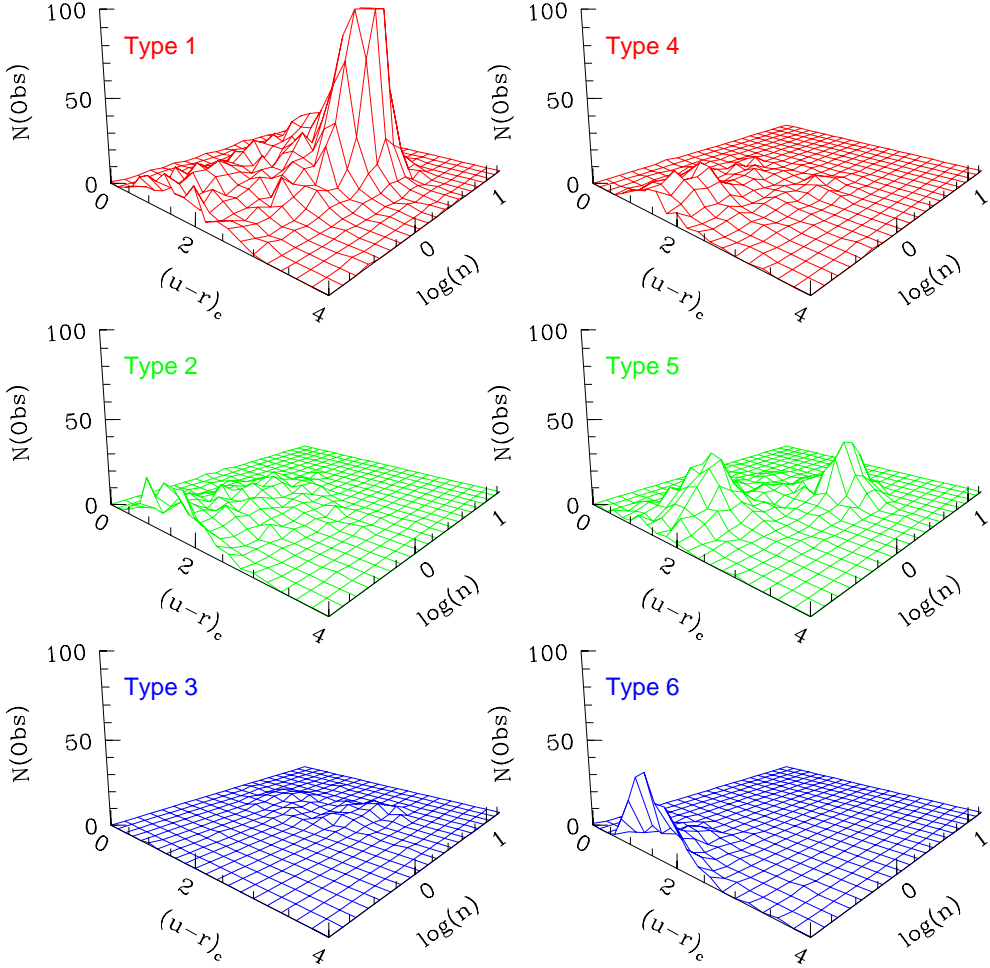}} 
\end{center}
\caption{The distribution of galaxy bulges (pre-filtering) in the 
$(u-r)_{{{\rm core}}}-\log(n)$ plane for each of the six profile types we 
identify in Section \ref{sec:interpret}.}
\label{fig:bulges}
\end{figure*}

\begin{figure*}
\begin{center}
{\leavevmode \epsfxsize=14.0cm \epsfysize=14.0cm \epsfbox{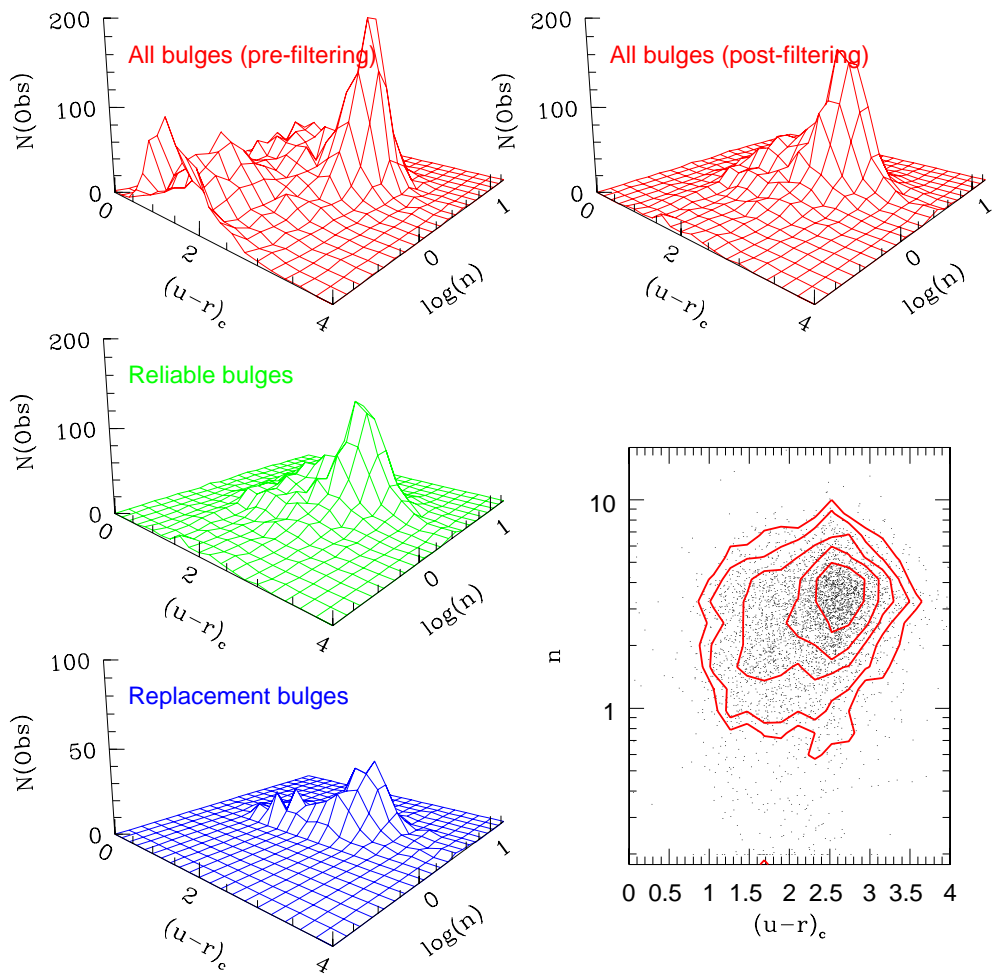}} 
\end{center}
\caption{Left column: the distribution of galaxy `bulges' in the 
$(u-r)_{{{\rm core}}}-\log(n)$ plane for: (top left) raw galaxy bulges 
(pre-filtering) of all types (see Figure \ref{fig:bulges}), (middle left) 
bulges from two-component \sersic\, + exponential model that pass through the 
logical filter unchanged, (bottom left) replacements from the \sersic-only 
catalogue that are interpreted as bulges. The right panels
show the distribution of bulges from the final filtered catalogue (reliable 
\sersic\,+ exponential bulges, plus replacement \sersic-only bulges) in 3D 
(top), and in 2D (bottom).}
\label{fig:bulges2}
\end{figure*}

\citet{driver_bimod} have demonstrated the existence of a bimodality in the 
colour-$\log(n)$ plane for luminous galaxies (when modelled with a 
single \sersic\, function); one can clearly identify a high-$n$ red peak, and 
a low-$n$ blue peak. The colours referred to here are the $(u-r)_{{{\rm core}}}$
PSF colours from SDSS photometry \citep[see][for more detail of our use of PSF 
colours]{driver_bimod}.
Furthermore, they suggest that the two peaks 
can be identified with bulges and discs, respectively. This differs in a subtle
but important way from the usual interpretation of the colour bimodality,
which is credited to early- and late-type galaxies rather than bulge and
disc stellar systems. Figure \ref{fig:bulges} shows the colour-$\log(n)$ plane 
for the \sersic\, (which one might initially interpret as a `bulge') component 
for each of the six two-component types we identify (see Figure 
\ref{fig:types}). The top-left panel of Figure \ref{fig:bulges2} shows the 
same cumulative raw distribution for all types 
together. This differs from \citet{driver_bimod} in that we have now separated 
the bulge and disc components, as Driver et al. argued should be done. It is 
clear from Figure \ref{fig:bulges} that most galaxies are `classical' type 1 
profiles, and of these, the majority of the `bulges' are associated with a 
high-$n$ red peak. However, for most galaxies of types 2--6, and for a 
small but significant fraction of type 1 galaxies, the bulges do not 
correspond to the high-$n$ red peak. To investigate this situation further we 
now consider each of the profile types in turn.

\begin{figure}
\begin{center}
{\leavevmode \epsfxsize=8.0cm \epsfysize=8.0cm \epsfbox{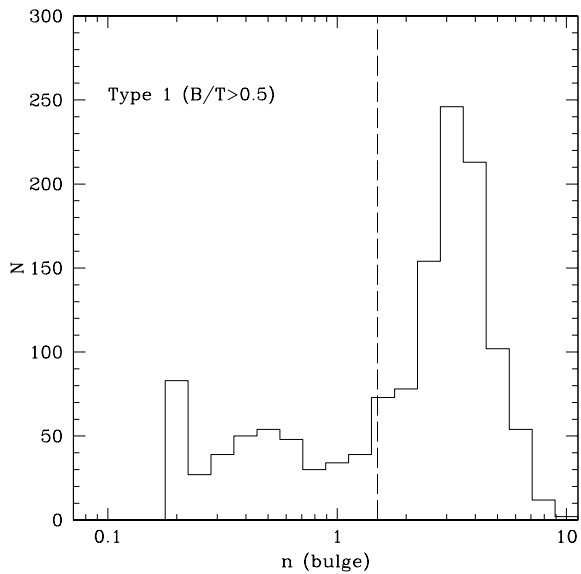}} 
\end{center}
\caption{The distribution of bulge Sersic index, $n$, for type 1 galaxies that 
are bulge dominated (i.e. $B/T>0.5$).}
\label{fig:type1n}
\end{figure}

The majority of the two-component fits correspond to a `classic' {{\bf type 1}}
profile. Analysis of the bulge properties of these galaxies suggests 
the presence of at least two populations in the majority of cases. The top 
left panel of Figure \ref{fig:bulges} shows that most objects lie in a clear 
peak around $n\sim3$, and PSF colour $(u-r)_{{{\rm core}}}\sim2.5$. These 
bulges appear to belong to plausible bulge-disc systems, and are accepted by 
us as correct decompositions, and therefore progress unchanged to the final 
filtered catalogue. Additionally, there is a tail of objects and a possible 
second peak around $n\sim0.5$, with a wider range in colours, albeit generally 
blue [$(u-r)_{{{\rm core}}}<2.0$]. This could be indicative of a population 
of pseudo-bulges or bars. Such a population might be expected in disc 
dominated systems ($B/T<0.5$), and these objects are accepted as having good 
decompositions and genuine bulges. An additional separation into bulges and 
pseudo-bulges/bars can be made at a later stage if required. 
However, pseudo-bulges are unlikely to be found in the high $B/T$ regime. 
Figure \ref{fig:type1n} shows the distribution in \sersic\, index for \sersic\,
dominated $(B/T>0.5)$ type 1 galaxies. Based on this distribution we accept 
objects with $n>1.5$ as having a good bulge-disc decomposition and they are 
included as such in the filtered catalogue, whilst type 1 objects with $n<1.5$ 
are replaced with single-component \sersic\, fits. These objects appear to be 
either pure disc systems or dwarf ellipticals. 

{{\bf Type 2}} galaxies are clearly disc-dominated objects (they all have 
$B/T<0.5$), with a possible small bulge. The addition of a weak \sersic\, 
function to an exponential function can be enough to perturb the total profile 
significantly (see the type 2 example in Figure \ref{fig:types}). However, the 
modelled perturbation to the total profile is not always at the centre, and 
is sometimes at much larger radii (analogous to the type 6 profile shown in 
Figure \ref{fig:types}, but without the profiles crossing). In these cases it 
would be incorrect to interpret the \sersic\, function as a bulge, and we 
therefore impose the additional restriction for type 2 objects that 
$R_{e}<0.5\times\,h$, ensuring that only \sersic\, functions modelling a 
central feature are considered as bulge-disc systems. We allow these objects 
to pass through to the filtered catalogue. Based on the colour-$\log(n)$ 
distribution in Figure \ref{fig:bulges}, the bulges of the majority 
of these galaxies do not correspond to the classical red, high-$n$ peak, and 
may be better interpreted as pseudo-bulges. A small fraction of these objects 
are red, but they are all highly inclined systems, and thus likely red due to 
dust. Those objects with $R_{e}>0.5\times\,h$ (i.e. the \sersic\, function 
does not correspond to a central perturbation) are replaced with 
single-component \sersic\, fits, and are interpreted as disc-only systems.

Only a small fraction of galaxies are classified as {{\bf type 3}}. These fits 
are erroneous and are sometimes caused by nearby neighbours (which GIM2D would 
be unaware of) contributing extra flux at large radii. They are generally
red, bulge-dominated systems, with high-$n$ and we choose to replace them with 
a \sersic-only fit, which will typically be later interpreted as a bulge-only
system. This class does contain a fraction ($<1$\%) of genuine bulge-disc 
systems that will be missed.

Many {{\bf type 4}} profiles can be considered true inversions, with the 
exponential function fitting a central feature, and the \sersic\, function 
fitting the outer disc. This can occur for a number of reasons. A PSF error 
can mean that the exponential is used to fit a sharp `spike' at the centre of 
the profile. 
Such spikes are always much smaller than the HWHM of the PSF. In other cases, 
the galaxy appears to be an irregular disc with active star formation. The size
and shape of the two components can be comparable, especially for the redder 
galaxies, when the modelled galaxy is often a dwarf. In all cases the 
bulge-disc interpretation would be erroneous and a single \sersic\, fit is 
preferred in the filtered catalogue. However a fraction of these galaxies
appear to be genuine bulge-disc systems with severe disc truncation. This
seems to be a real phenomenon, and is likely to be an increasing problem with 
future, deeper data sets. Similarly, at higher redshifts, discs are seen to
be described with a \sersic\, function requiring $n<1$ \citep{tamm}.
Unfortunately, modelling disc truncation is beyond the scope of GIM2D, and it 
is not possible to fit the 2D images accurately. Therefore, we choose to 
replace these fits with single-component \sersic\, functions, and accept that 
a small fraction ($<$5\%) of bulge-disc systems may be missed, or rather,
not decomposed.

The {{\bf type 5}} galaxies all have $B/T>0.6$, and are therefore dominated by 
the \sersic\, component. The addition of an exponential function (that does 
not cross) makes little difference to the total profile.
The distribution in \sersic\, index and colour is clearly 
bimodal (figure \ref{fig:bulges}), and there is a correlation 
between (two-component) \sersic-`bulge' parameters, and the \sersic-only 
parameters. These galaxies tend to be blue, low-$n$ disc-only systems, or 
red, high-$n$ bulge-only systems (i.e. spheroids). They are best fitted by a 
single \sersic\, component, and these fits are therefore used in the filtered 
catalogue. 

The fits to {{\bf type 6}} galaxies are clearly incorrect if interpreted as 
bulge-disc decompositions. From Figure \ref{fig:bulges} the `bulge' components 
have very low $n$, and blue $(u-r)$ colours. These galaxies also have low 
\sersic\, index ($n<1$) from the \sersic-only fits. They appear to be either 
dwarf galaxies or very flat ($n<1$) discs with active star formation and/or 
other irregularities. In all cases a single \sersic\, fit is adopted as more 
appropriate. 

Finally, those objects that only have one component from the 
\sersic\, + exponential model fits ({{\bf type 7}} and {{\bf type 8}}) are 
replaced with single-component \sersic-only fits.

\begin{figure}
\begin{center}
{\leavevmode \epsfxsize=8.0cm \epsfysize=8.0cm \epsfbox{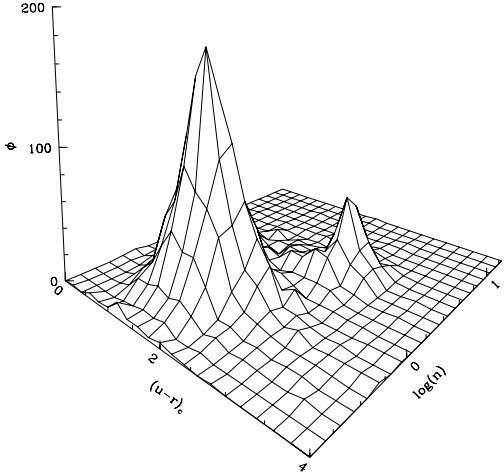}} 
\end{center}
\caption{The distribution of those galaxies replaced with \sersic-only fits by 
the logical filter in the $(u-r)_{{{\rm core}}}-\log(n)$ plane. We choose 
$n=1.5$ as the division between the two populations.}
\label{fig:replacements}
\end{figure}

\begin{figure*}
\begin{center}
{\leavevmode \epsfxsize=13.0cm \epsfysize=12.0cm \epsfbox{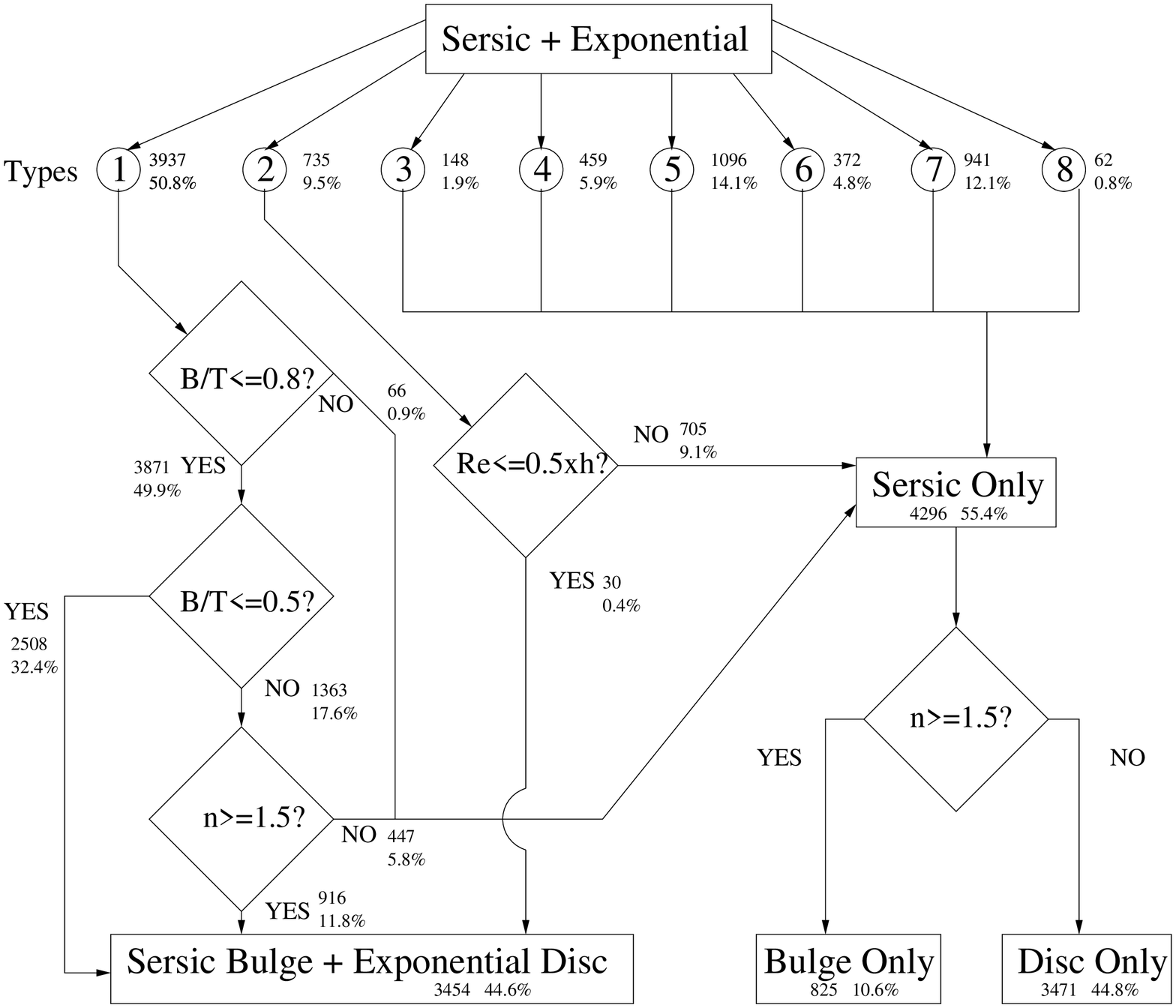}} 
\end{center}
\caption{A summary of the steps involved in the logical filter for the
\sersic+exponential catalogue.}
\label{fig:flow}
\end{figure*}

In summary, all profiles were replaced with \sersic-only fits except for types 
1 and 2, which generally keep the two-component bulge-disc decomposition. The 
exceptions are when (i) $B/T>0.8$, or (ii) a low-$n$ ($<1.5$) \sersic\, 
component dominates ($B/T>0.5$) for the type 1 galaxies, and (iii)
when $R_{e}>0.5\times\,h$ (i.e. the \sersic\, function models a non-central 
perturbation) for the type 2 galaxies. The logical filter is summarised in 
Figure \ref{fig:flow}. After the
filtering process, the full catalogue contains 3454 two-component objects where
the \sersic\, component is treated as a bulge, and the exponential component
is treated as a disc. The remaining galaxies in the catalogue (4296 objects)
are fitted with a single \sersic\, component. The distribution of these
\sersic-only objects in the colour-$n$ plane is shown in Figure 
\ref{fig:replacements}, where the bimodal nature is evident. Based on this 
distribution we choose $n=1.5$ as the division between the two populations. 
We classify  825 galaxies with $n>=1.5$ as bulge-only objects, and 3471 
galaxies with $n<1.5$ as discs. As already discussed, these replacements
will include a small fraction ($<6$\%) of type 3 or type 4 galaxies that are 
genuine bulge-disc systems, that are misinterpreted here as bulge-only or 
disc-only.

As a useful verification of this bulge- or disc-only interpretation 
we also looked at the above \sersic-only replacements in the 
$R^{1/4}$ + exponential catalogue. For the replacement `disc-only' objects
80.1\% of galaxies had a best-fitting $R^{1/4}$ + exponential model with 
$B/T=0$ (i.e. GIM2D found a single exponential disc the most optimal model). 
The remainder of the replacement discs were all in disc-dominated systems with 
$B/T<0.5$ (half of them with $B/T<0.1$). For the replacement `bulge-only' 
objects, the $R^{1/4}$ + exponential model produced mostly high $B/T$ output, 
with a median of $B/T\sim0.8$, suggesting that this population is indeed 
dominated by E/S0 galaxies.


\section{Repeat Observations and Parameter Accuracy}
\label{sec:repeat}

The geometry of the MGC is such that each pointing of the INT Wide Field Camera
overlaps by approximately 0.027 deg$^{2}$ with the previous one 
\citep[see Figure 2 in][]{liske}, building up a mosaiced strip across the 
sky. Objects that lie in these overlap regions are therefore observed twice. 
Although the MGC catalogues only utilise one of these observations, the 
imaging data exists and SExtractor catalogues, and GIM2D input images were 
prepared for all 702 duplicate galaxies. Here we remove those objects 
with incorrect masks (see Section \ref{sec:reanal}), reducing the number of
twice-observed comparison objects to 682.

In each case, repeat observations lie on a different CCD of the WFC, with 
different PSFs, airmasses, sky brightnesses, noise, and seeing conditions. 
Often, the observations were carried out on a different night or even as 
part of different observing runs several months or years apart. They therefore 
provide an excellent test of the repeatability of GIM2D fits for a diverse and
representative sample of galaxies under different conditions.

In all three catalogues acceptable repeatability (most measurements agree 
better that 20\%) is obtained if a bulge/disc 
component size limit is imposed such that the half-light radius is larger than 
the half-width half maximum of the seeing disc (half-light radius $>$ 
0.5$\times$seeing). Bulges are compared when $R_{e}>0.5\Gamma$, and discs are 
compared when $1.678h>0.5\Gamma$, where $\Gamma$ is the average seeing for the 
image the galaxy lies in. Below this threshold the differences between repeat 
measurements begin to increase. In addition we also find that the repeatability
is poor for less luminous components, and we therefore impose a $M=-17$ B mag 
limit on the component magnitudes. Note that the uncertainties discussed here 
are not included in the final structural catalogues which only contain the 
68\% confidence intervals computed by GIM2D (see Section \ref{sec:cats}).

Profiles from the repeat observations have been interpreted according to the 
rules described in  Section \ref{sec:logic}, and replacements made where 
necessary. In most cases the profile type is the same for each pair of 
observations of an object, but where the profile types differ, components 
interpreted as bulges or discs are directly compared (e.g. even if a galaxy 
has a type 2 profile in one measurement and a type 1 profile in the other, the 
bulge and disc components are still compared). When one measurement of an 
object has resulted in a single-component \sersic\, profile and the other 
measurement uses a two-component profile (e.g. type 1), the \sersic-only 
profile is interpreted as either a bulge or a disc (see Section 
\ref{sec:logic}), and compared with the corresponding component in the other 
fit. Type discrepancies such as this occur about 12\% of the time. For each 
parameter we compute the mean residual and the 3$-\sigma$ clipped standard 
deviation. In all cases the size of residuals correlates with the apparent
half-light radii of the components.

\begin{figure*}
\begin{center}
{\leavevmode \epsfxsize=15.0cm \epsfysize=15.0cm \epsfbox{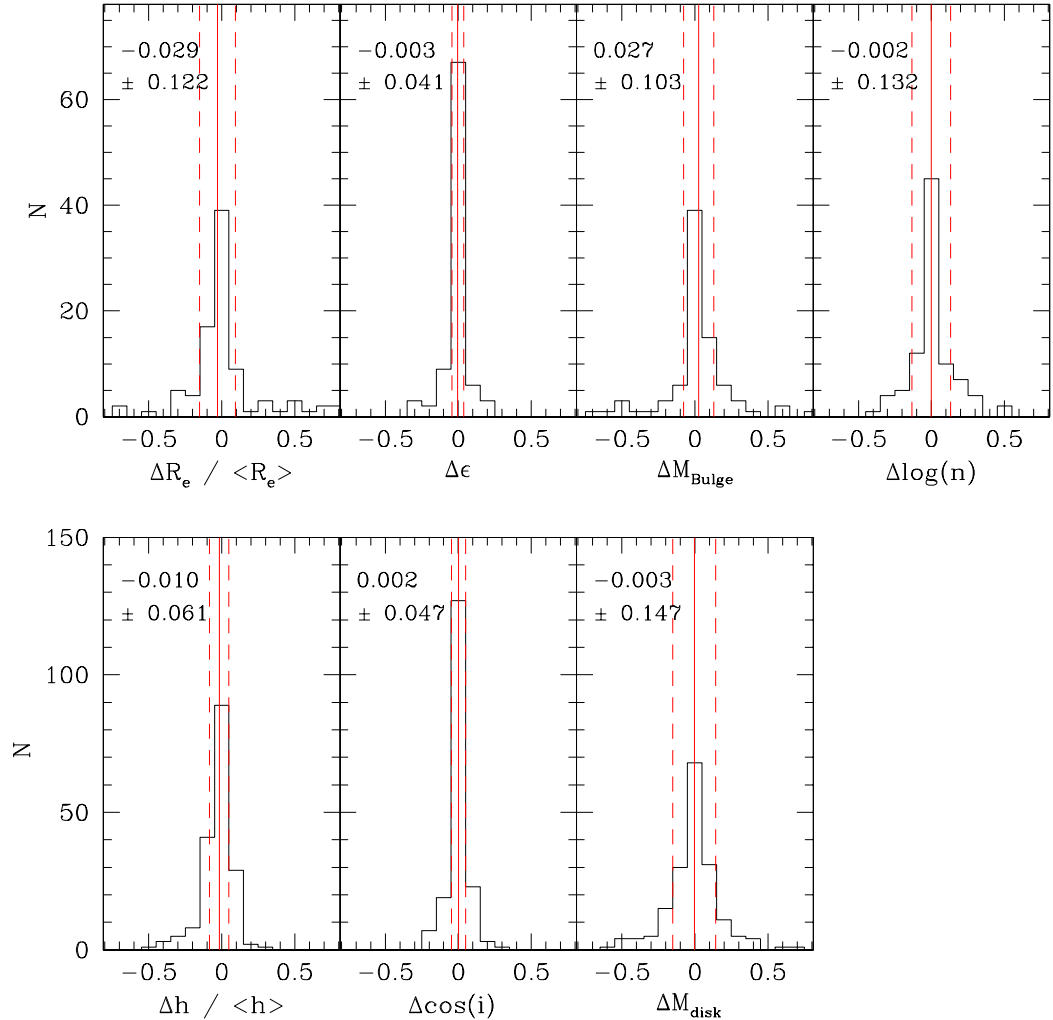}} 
\end{center}
\caption{Parameter comparison between repeat observations of galaxies. Bulge 
(top panel) and disc (lower panel) parameters are derived from \sersic-bulge/
exponential-disc filtered catalogues. The solid lines correspond to the mean
difference between repeat measurements and the dashed lines show the standard 
deviation. In each panel the mean and standard deviation is noted.}
\label{fig:secomp}
\end{figure*}

Figure \ref{fig:secomp} shows a comparison between bulge and disc parameters
from repeat observations for the filtered catalogue using 
two-component \sersic\,+exponential fits (or \sersic\, replacements). 
In all cases the mean is close to zero. We find bulge $R_{e}$ differences have
a standard deviation of $12.2\%$, the \sersic\, index, $n$, has a standard 
deviation of $\Delta \log(n) = 13.2\%$. Ellipticity differences have a standard
deviation of $4.1\%$, and the bulge magnitudes, $M_{{{\rm bulge}}}$, have a standard deviation of 0.10 B mag. For discs a larger sample can be used, as more 
discs are larger than the $0.5\Gamma$ cut. Scale lengths, $h$, have a standard 
deviation of 6.6\%, and disc inclinations, $\cos(i)$, have a standard deviation
of 4.7\%. Finally, disc magnitudes have a standard deviation of 0.15 B mag. The
largest residuals are mostly due to those objects that have 
different types between repeat observations (e.g. one observation results in 
a  high $B/T$ type 1 bugle-disc classification, and the other observation 
results in a type 5 bulge only classification in which cases only the bulge 
parameters are compared).

\begin{figure*}
\begin{center}
{\leavevmode \epsfxsize=15.0cm \epsfysize=15.0cm \epsfbox{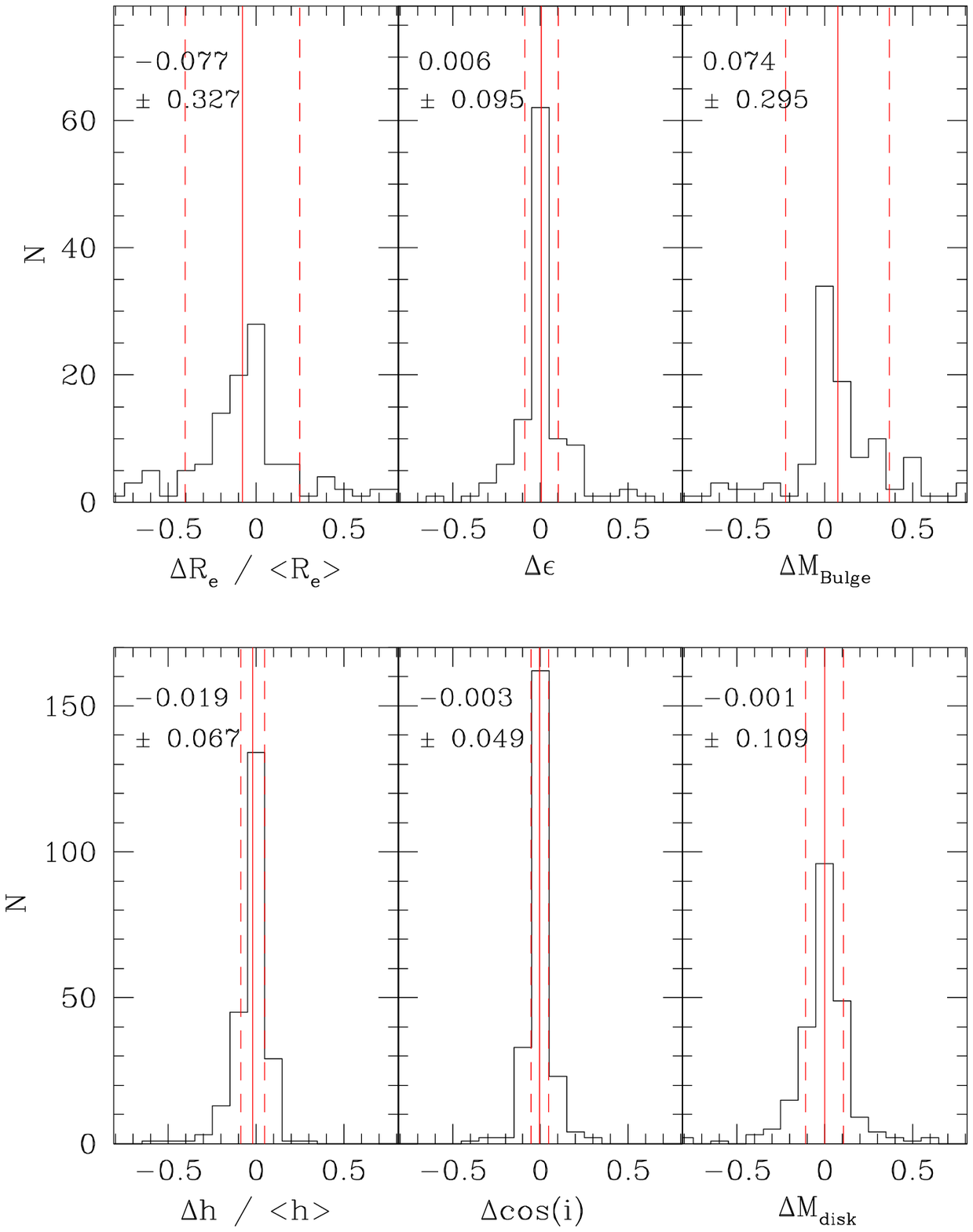}} 
\end{center}
\caption{Comparison between repeat observations of galaxies for bulge (top 
panel) and disc (lower panel) parameters derived from $R^{1/4}$ bulge + exponential disc filtered catalogues. The solid lines correspond to the mean
difference between repeat measurements and the dashed lines show the standard 
deviation. In each panel the mean and standard deviation is noted.}
\label{fig:dvecomp}
\end{figure*}

Figure \ref{fig:dvecomp} shows histograms of the distribution in differences
between parameters from repeat observations using the de Vaucouleurs + 
exponential catalogue.
Once again, the catalogues have been filtered according to the rules described 
in Section \ref{sec:interpret}. These measurements also show an acceptable 
level of repeatability, although the scatter in the bulge differences is larger
than that found using the \sersic\, + exponential catalogue. Bulge standard 
deviations are: $\Delta\,R_{e}/R_{e}=32.7\%$, 
$\Delta\,\epsilon/\epsilon=9.5\%$, and $\Delta\,M_{{{\rm bulge}}}=0.29$ mag. 
For discs, measurements are comparable to the \sersic\, + exponential 
catalogue, with standard deviations of $\Delta\,h/h=6.7\%$, 
$\Delta\,\cos(i)=4.9\%$ and $\Delta\,M_{{{\rm disc}}}=0.11$ mag.

\begin{figure*}
\begin{center}
{\leavevmode \epsfxsize=15.0cm \epsfysize=7.5cm \epsfbox{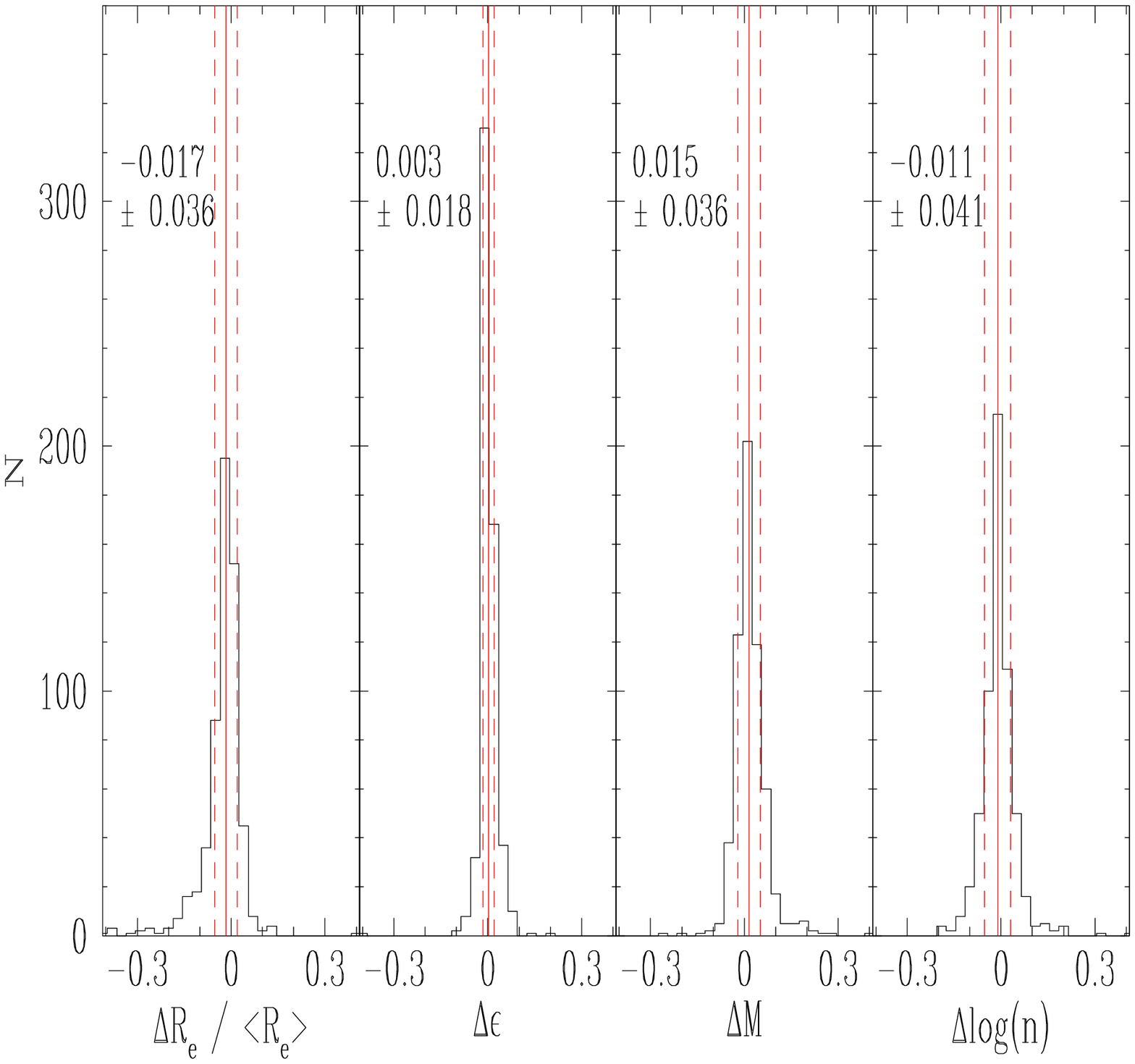}} 
\end{center}
\caption{Comparison between repeat observations of galaxies for parameters 
derived from single component \sersic\, fits. The solid lines correspond to the
mean difference between repeat measurements and the dashed lines show the 
standard deviation. In each panel the mean and standard deviation is noted.}
\label{fig:sersiccomp}
\end{figure*}

Residuals for measured parameters for single-component \sersic\, fits
are shown in Figure \ref{fig:sersiccomp}. As would be expected this much
simpler model produces much more repeatable results. In this case the 
standard deviations are $\Delta\,R_{e}/R_{e}=3.7\%$, 
$\Delta\,\epsilon/\epsilon=2.0\%$, $\Delta\,M=0.04$ mags, and 
$\Delta\,\log (n)= 4.2\%$. 

\section{Fundamental Bulge and Disc Properties}
\label{sec:bulgedisc}

\begin{figure}
\begin{center}
{\leavevmode \epsfxsize=8.0cm \epsfysize=8.0cm \epsfbox{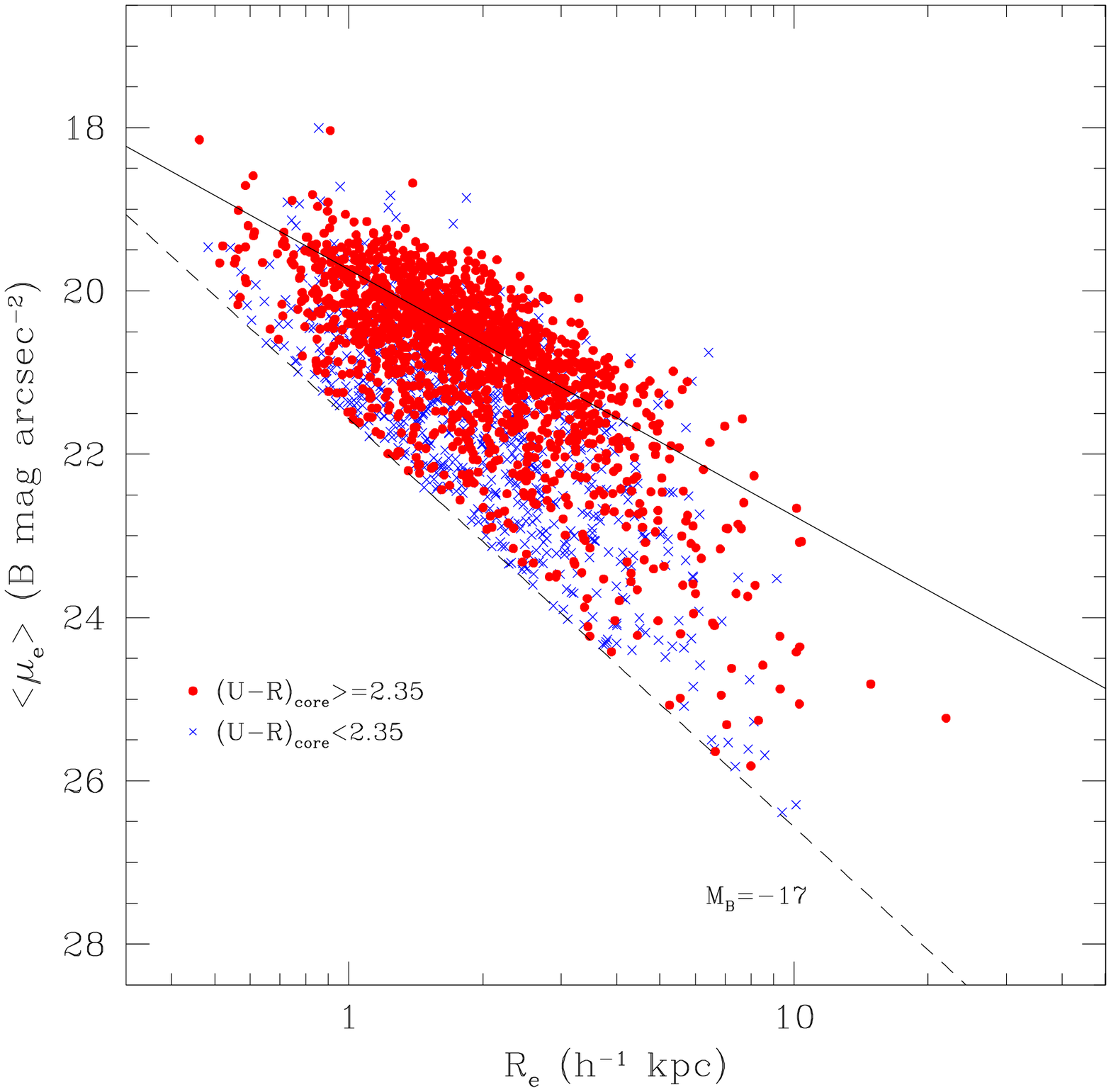}} 
\end{center}
\caption{The $\langle\mu_{e}\rangle-\log(R_{e})$ distribution for all MGC 
bulges. The red circles show objects with 
$(u-r)_{{{\rm core}}}\ge2.35$, and the blue crosses denote objects with 
$(u-r)_{{{\rm core}}}<2.35$. The dashed line shows the $M_{{{\rm bulge}}}=-17$
$B$ mag limit, and the solid line shows the \citet{kormendy} relation. The red 
population follows the Kormendy relation well, with most of the scatter coming
from objects with $B/T<0.5$. The blue (pseudo-)bulges do not appear 
to fit the relation.} 
\label{fig:korm}
\end{figure}

To demonstrate the consistency and suitability of the separation into bulge and disc 
populations we recover two fundamental statistical results: the Kormendy
relation \citep{kormendy} and the relation between disc central surface
brightness and disc scale-length \citep{dejong96b,graham2001}. In Figure 
\ref{fig:korm} the size-surface brightness relation is shown for {{\it all}} 
bulges with $R_{e}>0.5\Gamma$ and $M_{{{\rm bulge}}}<-17$ mag, along with the 
relation of \citet{kormendy}. The majority of red bulges follow the Kormendy 
relation well (especially for large values of $B/T$), however the smaller 
population of blue bulges do not. For reasons explained in 
\citet{graham_guzman}, faint spheroids do not follow the Kormendy relation, 
having smaller $R_{e}$, and fainter $\langle\mu_{e}\rangle$.

\begin{figure}
\begin{center}
{\leavevmode \epsfxsize=8.0cm \epsfysize=8.0cm \epsfbox{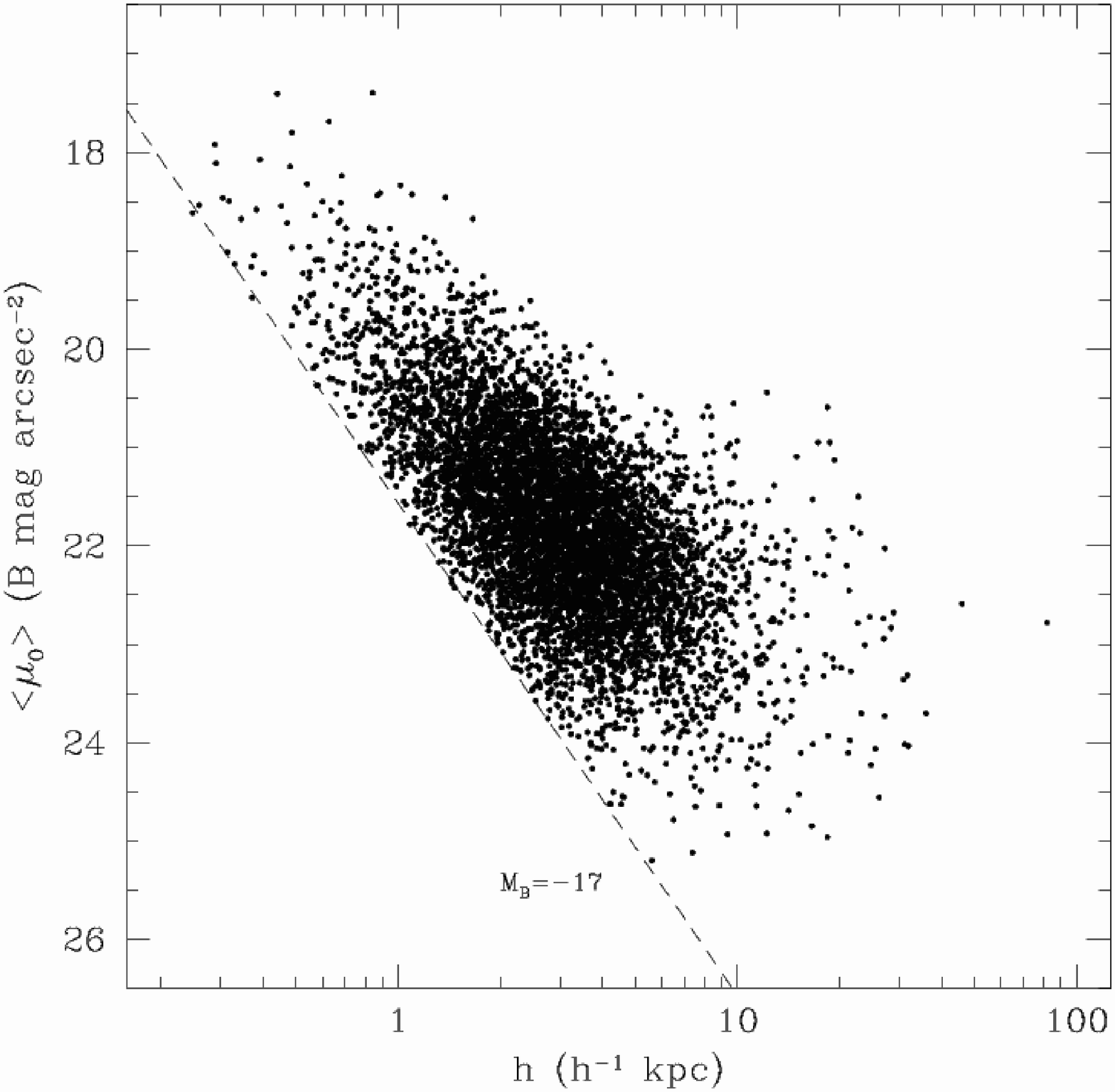}} 
\end{center}
\caption{The $\mu_{0}-\log(h)$ distribution 
for all reliable discs in the MGC (using the filtered \sersic+exponential 
catalogue). The dashed line shows the $M=-17$ limit. Discs with larger 
scale-lengths have fainter central surface brightnesses.}
\label{fig:freeman}
\end{figure}

Figure \ref{fig:freeman} shows the distribution in scale-length, $h$,
and central surface brightness, $\mu_{0}$, for discs from 
the \sersic+exponential catalogue after filtering (see Section 
\ref{sec:logic}). Only galaxies with $1.678h>0.5\Gamma$, and 
$M_{{{\rm disc}}}<-17$ mag (see Section \ref{sec:repeat}) are shown. 
The dashed line shows the $M_{{{\rm bulge}}}=-17\,B$ mag limit. It is clear
that larger discs have larger (i.e. fainter) values for their central
surface brightness as shown by other authors \citep{dejong96b,graham2001}.


\begin{figure}
\begin{center}
{\leavevmode \epsfxsize=8.0cm \epsfysize=8.0cm \epsfbox{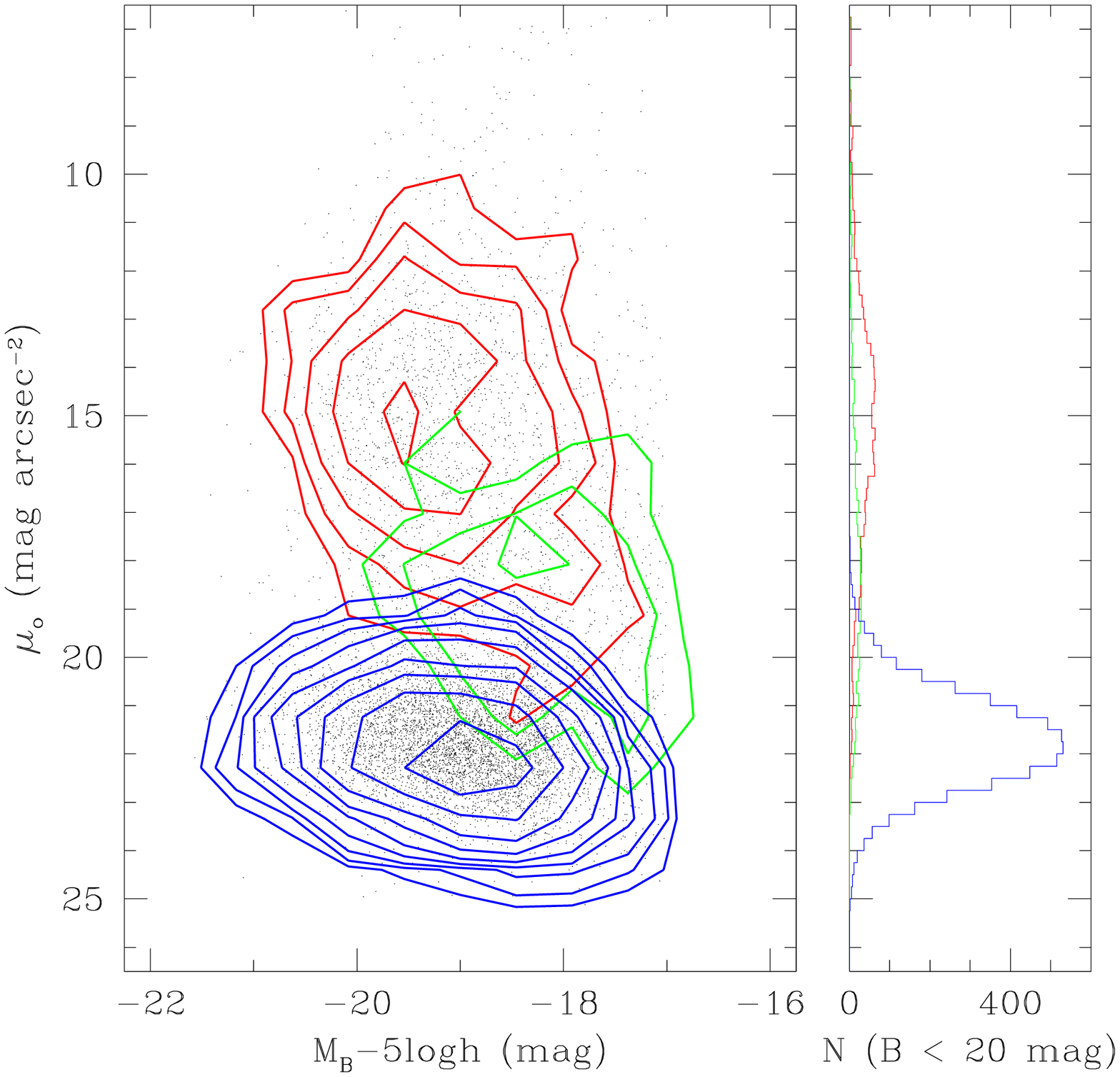}} 
\end{center}
\caption{The distribution of galaxy component properties in the 
$M_{B}-\mu_{0}$ plane is shown by the dots. Red contours outline the 
distribution of bulge components, and blue contours outline the distribution 
of disc components. Two clear populations are evident. The magenta contours 
outline the blue ($(u-r)_{{{\rm core}}}<2.35$) bulge components, which appear 
to lie in an intermediate position. The right panel shows histograms of the 
number of each type of component. The number of discs is boosted because they 
are more likely to meet the size criteria than bulges (or pseudo-bulges).}
\label{fig:mu0}
\end{figure}

The right panel of Figure \ref{fig:bulges2} shows the distribution of
bulges in the colour-$\log(n)$ plane from the \sersic+exponential catalogue
(post-filtering). This distribution can be compared with that presented in 
\citet{driver_bimod} who consider {{\it total}} rather than
{{\it component}} properties, and identify two distinct peaks. In Figure 
\ref{fig:bulges2}, the bulges clearly form a single red, high-$n$ peak, 
although there is a small shoulder of bluer, low-$n$ objects.
That the majority of bulges identify with the red, high-$n$ peak  
supports the conclusions of \citet{driver_bimod} that the colour bimodality
reported in the galaxy population is best explained by two fundamental
components: bulges and discs, rather than two different galaxy populations.
Furthermore, this probably reflects two dominant formation processes and 
epochs. The small number of blue, low-$n$ bulges may also be evidence of a 
third population of pseudo-bulges resulting from secular evolution.
The bimodal nature of the different components is again evident in Figure 
\ref{fig:mu0}, which shows the relationship between the model $\mu_{0}$, and
$M_{B}$ for bulge, disc, and pseudo-bulge components. It is clear that there 
are two distinct groupings with the high surface brightness bulges and the low 
surface brightness discs. The small number of blue, pseudo-bulges appear to 
have intermediate surface brightnesses.

\section{Conclusions}

We have used GIM2D to perform 2D model fits (including bulge-disc 
decomposition) for 10095 galaxies with $\bmgc<20$ mag from the Millennium 
Galaxy Catalogue. We initially produced three catalogues using: (1) a single-
component \sersic\, model, (2) a \sersic+exponential bulge-disc model, and
(3) an $R^{1/4}$+exponential bulge-disc model. We find that there is good 
agreement between GIM2D centroids, magnitudes, and half-light radii, with those
in the \citet{liske} MGC catalogues. However, there are a significant number of
objects that have incorrect mask images from SExtractor which had to be 
corrected manually. 

When a two-component model is used, we find that a significant 
fraction of galaxies have mathematically `good' fits, which may not be
the most appropriate or meaningful. The components may be inverted, cross
each other once, twice, or not at all. In addition, it is also important to 
stress that many galaxies are {{\it not}} two component systems, and simply 
interpreting one component as `bulge' and the other as `disc' would clearly be 
wrong. We identify eight distinct profile types based on which of the two 
components dominate at the galaxy centre, at $\mu=26$ mag arcsec$^{-2}$, and
whether the components cross. 

The majority ($\sim50\%$) of objects are `classical' type 1
galaxies where the \sersic\, function dominates at the centre (modelling a 
bulge), and the exponential function dominates the outer regions (modelling
a disc). The remaining objects often require more careful 
interpretation. The different types are sometimes due to phenomena such 
as disc truncation, multiple nuclei, or strong irregularities, all of which 
lie beyond the scope of the simple GIM2D model. Using the `type' scheme we 
implement a strategy to ensure only genuine bulge-disc systems are fitted with 
a \sersic+exponential model, and replace all other objects with 
single-component \sersic\, fits. In most cases only single component systems 
end up with \sersic-only fits. A small number ($<6\%$) of geniune 
bulge+disc systems are classified as type 3 or type 4 (especially those with 
disc truncation), and will therefore be mis-classified as single-component
systems. This `logical filtering' process results in a fourth catalogue, where 
each galaxy is treated as either bulge+disc, bulge-only, or disc-only, and it 
is this catalogue that we suggest contains the most appropriate, physically 
meaningful modelling of the galaxy population. All five structural catalogues 
are made publically available with this paper (see Appendix A). 
 
To test the repeatability and accuracy of our measurements, we applied GIM2D 
twice to the objects that lie in the overlap regions between the
individual pointings that make up the MGC strip. We find that if size
and absolute magnitude cuts are imposed, parameter measurements are repeatable 
at less than the $15\%$ level. Although the catalogues contain entries for all 
objects, we recommend imposing cuts in size (half-light radius $>0.5\Gamma$) 
and component absolute magnitude ($M<-17\,B $ mag) in order to obtain reliable 
measurements. To demonstrate the accuracy of the final, logically filtered
catalogue we recover the fundamental Kormendy and $\mu_{0}-\log(h)$ relations for
bulges and discs respectively. In a series of future papers, the catalogues 
presented here will be used in detail to measure bulge and disc luminosity 
functions, bivariate brightness distributions, and size distributions. We are
also using the catalogues to measure the supermassive black hole mass function 
and to study the effects of disc opacity.

Finally, we have used the catalogues to show that the galaxy colour-$\log(n)$ 
bimodality is due to the two-component nature of galaxies, rather than two
distinct galaxy populations. Luminous bulges generally occupy the red, 
high-$n$ peak, and discs occupy the blue, low-$n$ peak. There is also some 
evidence for a third population of blue, pseudo-bulges. The observed bimodality
may be the end-result of two distinct processes and timescales associated with 
bulge and disc formation. We conclude that routine bulge-disc decomposition is 
essential at all redshifts to fully understand the evolution of the luminous 
($M_{B}<-17$ mag) galaxy population. Three major steps forward are now 
required: (1) high-resolution, deep near-IR survey data, (2) the expansion
of bulge-disc decomposition software to accommodate real phenomena such as 
nuclei and disc truncation, and (3) deeper studies to characterise the
dwarf ($M_{B}>-17$ mag) population.

\section*{Acknowledgements}
We thank Luc Simard for making his GIM2D code publically available.
The Millennium Galaxy Catalogue consists of imaging data from the Isaac Newton
Telescope obtained through the ING Wide field Camera Survey Programme. The
Isaac Newton Telescope is operated on the island of La Palma by the Isaac 
Newton Group in the Spanish Observatorio del Roque de los Muchachos of the
Instituto de Astrof\'{i}sica de Canarias. Spectroscpic data comes from the 
Anglo-Australian Telescope, The ANU 2.3--m, the ESO New Technology Telescope, 
the Telescopio Nazionale Galileo, and the Gemini North Telescope. The 
survey has been supported through grants from the Particle Physics and 
Astronomy Research Council (UK), and the Australian Research Council (AUS). 
The data and data products are publically available from 
http://www.eso.org/$\sim$jliske/mgc/ or on request from J. Liske.

\bibliographystyle{./mn}
\bibliography{./paul_references}

\onecolumn

\appendix

\section{Catalogues and catalogue parameters} We provide five
structural catalogues each containing 118 parameters for the 10 095
galaxies assembled from MGC-BRIGHT. The catalogues are available
from the MGC website at http://www.eso.org/$\sim$jliske/mgc as:

\begin{tabular}{ll} \\
master.bsersic.cat & -- Catalogue containing S\'ersic-only profile fits with $e < 0.7$. \\
master.dsersic.cat & -- Catalogue containing S\'ersic-only profile fits with $i < 85.0$. \\
master.dve.cat & -- Catalogue containing de Vaucouleur plus exponential profile fits. \\
master.se.cat & -- Catalogue containing S\'ersic plus exponential profile fits. \\
master.logic.cat & -- Bulge and disc catalogue after processing through the logical filter, containing \sersic+exponential bulge+disc profiles, and bulge-only
or disc-only single component \sersic\, profiles.\\
\end{tabular}

~

\noindent
The parameter file relevant for all five data files is:

\begin{tabular}{ll} \\
master.par & -- List of parameters included in the above five files. \\
\end{tabular}

~

\noindent
Full details for each column entry are given below.

~

\noindent {\it Column 1:} (ID) MGC unique identification number as listed
in NED numbers 00000-69999 represent original SExtractor detections
and numbers 90000+ represent sources that have been rebuilt
\citep[see][]{liske}.

\noindent {\it Column 2:} ($\bmgc$) The SExtractor {\sc BEST}
magnitude corrected for Galactic extinction via the \citet{schlegel}
dust maps, the photometric system is defined in \citet{liske} and
conversions to various filters are listed in the appendix of
\citet{cross}.

\noindent {\it Column 3:} ($\langle \mu_e \rangle$) The apparent effective $\bmgc$ surface
brightness inside the empirically measured seeing corrected half-light
radius: $\langle \mu_e \rangle = \bmgc +2.5 \log_{10}(2 \pi [R_e^o]^2)$ (mag arcsec$^{-2}$).

\noindent {\it Column 4 \& 5:} (RA \& DEC) Right ascension and declination in J2000.0 (deg).

\noindent {\it Column 6:} (CLASS) Classification parameter: 1=galaxy, 8=star.

\noindent {\it Column 7:} ($R_e$) The empirically measured
half-light radius using the positional angle and ellipticity provided
by SExtractor. The magnitudes are assumed total (arcsec).

\noindent {\it Column 8:} ($R_e^o$) Seeing corrected half-light radius ,i.e.,
$R_e^o=\sqrt{R_e^2 - 0.32 \Gamma^2}$ \citep[see][]{driver05}.
\label{lastpage} (arcsec).

\noindent {\it Column 9:} ($\Gamma$) Full width at half maximum of the
seeing (arcsec).

\noindent {\it Column 10:} (spec) Best spectral fitting template from
\citet{poggianti} \citep[see][for fitting details]{driver05}.

\noindent {\it Column 11:} ($k_z(B)$) $\bmgc$ K-correction (mag).

\noindent {\it Column 12:} ($z$) Best redshift for this galaxy.

\noindent {\it Column 13:} ($Q_z$) Redshift quality flag: 
1=targeted but no redshift, 2= tentative redshift
measurement, 3= reliable redshift, 4 = definite redshift, 5=
unequivocal redshift 9 = not targeted.

\noindent {\it Column 14:} ($u_{SDSS}$[Pet]) SDSS-DR1 extinction corrected Petrosian apparent magnitude (AB mag). 

\noindent {\it Column 15:} ($g_{SDSS}$[Pet]) SDSS-DR1 extinction corrected Petrosian apparent magnitude (AB mag).

\noindent {\it Column 16:} ($r_{SDSS}$[Pet]) SDSS-DR1 extinction corrected Petrosian apparent magnitude (AB mag).

\noindent {\it Column 17:} ($i_{SDSS}$[Pet]) SDSS-DR1 extinction corrected Petrosian apparent magnitude (AB mag).

\noindent {\it Column 18:} ($z_{SDSS}$[Pet]) SDSS-DR1 extinction corrected Petrosian apparent magnitude (AB mag).

\noindent {\it Column 19:} (ID) Duplicate of column 1.

\noindent {\it Column 20:} ($M_{u_{SDSS}}$[Pet]) SDSS-DR1 extinction corrected Petrosian Absolute$^1$ magnitude (AB mag).

\noindent {\it Column 21:} ($M_{\bmgc}$[Kron]) MGC extinction corrected Kron Absolute$^1$ magnitude ($\bmgc$ mag).

\noindent {\it Column 22:} ($M_{g_{SDSS}}$[Pet]) SDSS-DR1 extinction corrected Petrosian Absolute$^1$ magnitude (AB mag).

\noindent {\it Column 23:} ($M_{r_{SDSS}}$[Pet]) SDSS-DR1 extinction corrected Petrosian Absolute$^1$ magnitude (AB mag).

\noindent {\it Column 24:} ($M_{i_{SDSS}}$[Pet]) SDSS-DR1 extinction corrected Petrosian Absolute$^1$ magnitude (AB mag).

\noindent {\it Column 25:} ($M_{z_{SDSS}}$[Pet]) SDSS-DR1 extinction corrected Petrosian Absolute$^1$ magnitude (AB mag).

\noindent {\it Column 26:} ($\mu_{u_{SDSS}}$[Pet]) SDSS-DR1 extinction corrected Petrosian Absolute$^1$ effective surface brightness (AB mag arcsec$^{-2}$).

\noindent {\it Column 27:} ($\mu_{\bmgc}$[Kron]) MGC extinction corrected Kron Absolute$^1$ effective surface brightness ($\bmgc$ mag arcsec$^{-2}$).

\noindent {\it Column 28:} ($\mu_{g_{SDSS}}$[Pet]) SDSS-DR1 extinction corrected Petrosian Absolute$^1$ effective surface brightness (AB mag arcsec$^{-2}$).

\noindent {\it Column 29:} ($\mu_{r_{SDSS}}$[Pet]) SDSS-DR1 extinction corrected Petrosian Absolute$^1$ effective surface brightness (AB mag arcsec$^{-2}$).

\noindent {\it Column 30:} ($\mu_{i_{SDSS}}$[Pet]) SDSS-DR1 extinction corrected Petrosian Absolute$^1$ effective surface brightness (AB mag arcsec$^{-2}$).

\noindent {\it Column 31:} ($\mu_{z_{SDSS}}$[Pet]) SDSS-DR1 extinction corrected Petrosian Absolute$^1$ effective surface brightness (AB mag arcsec$^{-2}$).

\noindent {\it Column 32:} ($M_{u_{SDSS}}$[PSF]) SDSS-DR1 extinction corrected PSF Absolute$^1$ magnitude (AB mag).

\noindent {\it Column 33:} ($M_{g_{SDSS}}$[PSF]) SDSS-DR1 extinction corrected PSF Absolute$^1$ magnitude (AB mag).

\noindent {\it Column 34:} ($M_{r_{SDSS}}$[PSF]) SDSS-DR1 extinction corrected PSF Absolute$^1$ magnitude (AB mag).

\noindent {\it Column 35:} ($M_{i_{SDSS}}$[PSF]) SDSS-DR1 extinction corrected PSF Absolute$^1$ magnitude (AB mag).

\noindent {\it Column 36:} ($M_{z_{SDSS}}$[PSF]) SDSS-DR1 extinction corrected PSF Absolute$^1$ magnitude (AB mag).

\noindent {\it Column 37:} ($\theta_1$) Angular size of 1h$^{-1}$pc at object
redshift: $\theta_1 = 3600\tan^{-1}$$[\frac{(1+z)}{1000 d_p}]$, a
value of 0.0 implies no redshift (arcsec).

\noindent {\it Column 38:} ($d_p$) Proper (co-moving) distance to the object ($h^{-1}$ Mpc).

\noindent {\it Column 39:} (c type) Continuum type: 1=El 15 Gyr, 2= Sa 7.4 Gyr, 3= Sc 2.2 Gyr \citep[see][]{driver_bimod}.

\noindent {\it Column 40:} (m type) Eyeball morphological type: 0 =
not classified, 1 = E/S0, 2= Sabc, 3=Sd/Irr (all $\bmgc < 19$ mag have been classified).

\noindent {\it Column 41:} (ID) Duplicate of column 1.

\noindent {\it Column 42:} (2dFGRS No.) Matched 2dFGRS serial number for this object (000000 if no match).

\noindent {\it Column 43:} ($\eta$) 2dFGRS $\eta$ parameter if matched \citep[see][-99.9 if no match]{madgwick}.

\noindent {\it Column 44:} (ID) Duplicate of column 1.

\noindent {\it Column 45:}    (L\_TOT)        Total flux (digital units).

\noindent {\it Column 46:}    (L\_TOT-)       Total flux error (-). 

\noindent {\it Column 47:}    (L\_TOT+)       Total flux error (+).  

\noindent {\it Column 48:}    (BULGE\_FRAC)   Bulge fraction (0=pure disc).

\noindent {\it Column 49:}    (BULGE\_FRAC-) Bulge fraction error (-).

\noindent {\it Column 50:}    (BULGE\_FRAC+)  Bulge fraction error (+).

\noindent {\it Column 51:}    (BULGE\_RE)     Bulge effective radius (pixels).

\noindent {\it Column 52:}    (BULGE\_RE-)    Bulge effective radius error (-).

\noindent {\it Column 53:}    (BULGE\_RE+)    Bulge effective radius error (+).

\noindent {\it Column 54:}    (BULGE\_E)      Bulge ellipticity.

\noindent {\it Column 55:}    (BULGE\_E-)     Bulge ellipticity error (-).

\noindent {\it Column 56:}    (BULGE\_E+)     Bulge ellipticity error (+).

\noindent {\it Column 57:}    (BULGE\_PA)     Bulge position angle. 

\noindent {\it Column 58:}    (BULGE\_PA-)    Bulge position angle error (-).

\noindent {\it Column 59:}    (BULGE\_PA+)    Bulge position angle error (+).

\noindent {\it Column 60:}    (R\_D)          Exponential disc scale length (pixels).

\noindent {\it Column 61:}    (R\_D-)         Exponential disc scale length error (-).

\noindent {\it Column 62:}    (R\_D+)         Exponential disc scale length error (+).

\noindent {\it Column 63:}    (DISC\_I)       Disc inclination (0=face-on).

\noindent {\it Column 64:}    (DISC\_I-)      Disc inclination error (-).

\noindent {\it Column 65:}    (DISC\_I+)      Disc inclination error (+).   

\noindent {\it Column 66:}    (DISC\_PA)      Disc position angle.

\noindent {\it Column 67:}    (DISC\_PA-)     Disc position angle error (-).

\noindent {\it Column 68:}    (DISC\_PA+)     Disc position angle error (+).

\noindent {\it Column 69:}    (X\_OFF)        X offset of galaxy centre (pixels).

\noindent {\it Column 70:}    (X\_OFF)        X offset error (-).

\noindent {\it Column 71:}    (X\_OFF)        X offset error (+).

\noindent {\it Column 72:}    (Y\_OFF)        Y offset of galaxy centre (pixels).

\noindent {\it Column 73:}    (Y\_OFF)        Y offset error (-).

\noindent {\it Column 74:}    (Y\_OFF)        Y offset error (+).

\noindent {\it Column 75:}    (BACK)         Background level (digital units).

\noindent {\it Column 76:}    (BACK-)        Background level error (-).

\noindent {\it Column 77:}    (BACK+)        Background level error (+).

\noindent {\it Column 78:}    ($n$)       Sersic index (=4 for de Vaucouleurs profile).

\noindent {\it Column 79:}    ($n$-)      Sersic index error (-).

\noindent {\it Column 80:}    ($n$+)      Sersic index error (+).

\noindent {\it Column 81:}    (CHI)          $\chi^2$ of GIM2D fit.

\noindent {\it Column 82:}    (RHALF)        GIM2D HLR (pixels).

\noindent {\it Column 83:}    (C1)           Concentration index alpha=1.

\noindent {\it Column 84:}    (A1)           Asymmetry index (background corrected) alpha=1.

\noindent {\it Column 85:}    (B1)           Background correction applied to A1.

\noindent {\it Column 86:}    (C2)           Concentration index alpha=2.

\noindent {\it Column 87:}    (A2)           Asymmetry index (background corrected) alpha=2.

\noindent {\it Column 88:}    (B2)           Background correction applied to A2.

\noindent {\it Column 89:}    (C3)           Concentration index alpha=3.

\noindent {\it Column 90:}    (A3)           Asymmetry index (background corrected) alpha=3.

\noindent {\it Column 91:}    (B3)           Background correction applied to A3.

\noindent {\it Column 92:}    (C4)           Concentration index alpha=3.

\noindent {\it Column 93:}    (A4)           Asymmetry index (background corrected) alpha=3.

\noindent {\it Column 94:}    (B4)           Background correction applied to A3.
                
\noindent {\it Column 95:} ($M_{\bmgc}$[GIM2D]) Total extinction corrected absolute$^2$ magnitude derived from L\_TOT (mag).

\noindent {\it Column 96:} ($M_{\bmgc}$(Bulge)[GIM2D]) Total extinction corrected absolute$^2$ magnitude derived from BULGE\_FRAC*L\_TOT (mag).

\noindent {\it Column 97:} ($M_{\bmgc}$(Disc)[GIM2D]) Total extinction corrected absolute$^2$ magnitude derived from (1-BULGE\_FRAC)*L\_TOT (mag).

\noindent {\it Column 98:} ($R_e$(Bulge)[GIM2D]) Half-light radius of bulge component: $R_e=0.333 $BULGE\_E$/\theta$ ($h^{-1}$ kpc). 

\noindent {\it Column 99:} ($\alpha_D$(Disc)[GIM2D]) Scale-length of disc component: $\alpha_D=0.333 $R\_D$/\theta$ ($h^{-1}$ kpc).

\noindent {\it Column 100:} ($\mu_0$ (Bulge)) Absolute central surface brightness: $\mu_o(\mbox{Bulge}) = \mu_e(\mbox{Bulge})-1.0857b$ \citep[see][eqn.~7]{graham_driver} (mag arcsec$^{-2}$).

\noindent {\it Column 101:} ($\mu_e$ (Bulge)) Absolute effective surface brightness at $R_e$ (Bulge): $\mu_e (\mbox{Bulge}) = \langle \mu_e \rangle (\mbox{Bulge}) +2.5 \log_{10} [\frac{ne^b}{b^{2n}}\Gamma(2n)]$ \citep[see][eqn.~9]{graham_driver} (mag arcsec$^{-2}$).

\noindent {\it Column 102:} ($\langle \mu_e \rangle$ (Bulge)) Absolute effective surface brightness of bulge component: $\langle \mu_e \rangle (\mbox{Bulge})= M_{\bmgc}(\mbox{Bulge}) + 2.5 \log_{10} (2 \pi R_e^2)+36.57$ \citep[see][eqn.~12]{graham_driver} (mag arcsec$^{-2}$).

\noindent {\it Column 103:} ($\mu_0$ (Disc)) Absolute central surface brightness: $\mu_o(\mbox{Disc}) = \mu_e(\mbox{Disc})-1.0857b$ \citep[see][eqn.~7]{graham_driver} (mag arcsec$^{-2}$).

\noindent {\it Column 104:} ($\mu_e$ (Disc)) Absolute effective surface brightness at $1.678 \alpha_D$: $\mu_e (\mbox{Disc}) = \langle \mu_e \rangle (\mbox{Disc}) +2.5 \log_{10} [\frac{ne^b}{b^{2n}}\Gamma(2n)]$ \citep[see][eqn.~9]{graham_driver} (mag arcsec$^{-2}$).

\noindent {\it Column 105:} ($\langle \mu_e \rangle$ (Disc)) Absolute effective surface brightness of disc component: $\langle \mu_e \rangle (\mbox{Disc})= M_{\bmgc}(\mbox{Disc}) + 2.5 \log_{10} (2 \pi (1.678 \alpha_D)^2)+36.57$ \citep[see][eqn.~12]{graham_driver} (mag arcsec$^{-2}$).

\noindent {\it Column 106:} ($R_x$) Radius at which the surface brightness of the disc equals the surface brightness of the bulge, i.e., 
$\langle \mu_e \rangle (\mbox{Bulge})+1.0857 b (\frac{R_x}{R_e(\mbox{Bulge})})^{\frac{1}{n}}=\langle \mu_e \rangle (\mbox{Disc}) +1.0857(\frac{R_x}{\alpha_D})$ \citep[see][eqn.~17]{graham_driver} (arcsec).

\noindent {\it Column 107:} ($\mu_x$) Surface brightness at $R_x$ (mag arcsec${-2}$).

\noindent {\it Column 108:} (p type) Profile type as defined in Section \ref{sec:interpret}.

\noindent {\it Column 109:} ($b$) the value which satisfies $\Gamma(2n)=2\gamma(2n,b)$ \citep[see][]{graham_driver}.

\noindent {\it Column 110:} (ID) Duplicate of column 1.

\noindent {\it Column 111:} ($(u-r)_g$) Rest frame global colour: $M_{u} (\mbox{Pet})-M_{r} (\mbox{Pet})$ (AB mag).

\noindent {\it Column 112:} ($(u-r)_c$) Rest frame core/bulge colour: $M_{u}(\mbox{PSF})-M_{r} (\mbox{PSF})$ (AB mag).

\noindent {\it Column 113:} ($(u-r)_d$) Rest frame outer/disc colour: $(u-r)_d = -2.5\log_{10}[\frac{ 10^{-0.4(u-r)_g}-\frac{B}{T}10^{-0.4(u-r)_c}}{1-\frac{B}{T}}]$ (AB mag).

\noindent {\it Column 114:} (ID) Duplicate of column 1.

\noindent {\it Column 115:} (FLAGS) SExtractor flags from original detection.

\noindent {\it Column 116:} (ID) Duplicate of column 1.

\noindent {\it Column 117:} ($S$) Galaxy significance: i.e., $\frac{ \phi(M,\langle \mu_e \rangle)}{N(M, \langle \mu_e \rangle)}$ \citep[see][]{driver05,driver_bimod}.

\noindent {\it Column 118:} ($\Delta S$) Error in $S$ \citep[see][]{driver05,driver_bimod}.

~

\noindent
Notes:

~

\noindent
$^1$ Absolute magnitudes are derived using $\Omega_M=0.3,
\Omega_{\Lambda}=0.7, H_0=100$ km s$^{-1}$ Mpc$^{-1}$, the appropriate
k-correction derived for each filter and an evolutionary correction of
the form $L_{z=0.0}=L_z(1+z)^{-\beta}$, where
$\beta=1.5,0.75,0.5,0.375,0.3,0.2$ for $u,\bmgc,g,r,i,z$ respectively.
Galaxies without redshifts have values of -99.9 in these columns.

\noindent
$^2$ GIM2D absolute magnitudes are derived using: $M_{\mbox{T,B,D}} =
Z_p -2.5 \log_{10}(X_{\mbox{T,B,D}}\mbox{L\_TOT})-(B-M_B)-A_B$ where
$X=1$, BULGE\_FRAC or (1-BULGE\_FRAC) for Total (T), Bulge (B) or Disc
(D) respectively (null values are set to 99.99).

\end{document}